\documentclass[prd,aps,floats,showpacs,twocolumn,twoside,preprintnumbers,
superscriptaddress,floatfix]{revtex4}
\usepackage{bm}
\usepackage[T1]{fontenc}
\usepackage[latin1]{inputenc}
\usepackage{amsmath}
\usepackage{amssymb}
\usepackage{mathrsfs}
\usepackage{graphicx,pstricks,epsfig}

\begin{document}

\preprint{\vbox{\hbox{Preprint no.}}}

\title{Neutrino emissivities and bulk viscosity in neutral two-flavor quark matter}
\author{J. Berdermann}
\email{jens.berdermann@dlr.de}
\affiliation{Deutsches Zentrum f{\"u}r Luft- und Raumfahrt (DLR), DE-17235 Neustrelitz, Germany}
\author{D. Blaschke}
\email{david.blaschke@ift.uni.wroc.pl}
\affiliation{Institute of Theoretical Physics, Uniwersity of Wroclaw, 50-204 Wroclaw, Poland}
\affiliation{
Bogoliubov  Laboratory of Theoretical Physics,
Joint Institute for Nuclear Research,
141980 Dubna,
Russia}
\affiliation{National Research Nuclear University (MEPhI), 115409 Moscow, Russia}
\author{T. Fischer}
\email{tobias.fischer@ift.uni.wroc.pl}
\affiliation{Institute of Theoretical Physics, Uniwersity of Wroclaw, 50-204 Wroclaw, Poland}
\author{A. Kachanovich}
\email{akachanovich@ift.uni.wroc.pl}
\affiliation{Institute of Theoretical Physics, Uniwersity of Wroclaw, 50-204 Wroclaw, Poland}

\begin{abstract}    
We study thermodynamic and transport properties for the isotropic
color-spin-locking (iso-CSL) phase of two-flavor superconducting quark matter under
compact star constraints within a NJL-type chiral quark model.
Chiral symmetry breaking and the phase transition to superconducting
quark matter leads to a density dependent change of quark masses,
chemical potentials and diquark gap. A self-consistent treatment of these
physical quantities influences on the microscopic calculations of transport
properties.
We present results for the iso-CSL direct URCA emissivities and
bulk viscosities, which fulfill the constraints on quark matter derived from
cooling and rotational evolution of compact stars. 
We compare our results with the phenomenologically successful, but yet  heuristic 2SC+X phase.
We show that the microscopically founded iso-CSL phase can replace the purely 
phenomenological 2SC+X phase in modern simulations of the cooling evolution 
for compact stars with color superconducting quark matter interior.
\end{abstract}


\maketitle


\section{Introduction}

Present astrophysical observational programmes monitoring compact stars (CS) have provided new, high-quality data for their static properties, thermal and spin evolution. These modern measurements constrain the equation of state (EoS) and the transport properties of dense matter in CS interiors \cite{Klahn:2006ir}, for a recent review see~\cite{Blaschke:2016}. In particular, the evidence for high masses \cite{Demorest:2010bx,Antoniadis:2013,Fonseca:2016tux} and large radii \cite{Bogdanov:2012md} of CSs, suggests that the EoS at high densities must be sufficiently stiff. This prompts the question for the possibility of deconfined quark matter at CSs interiors \cite{Ozel:2006km}. In this debate, it has been demonstrated that microscopic models of quark matter EoS allow for extended quark cores of CS, while satisfying current mass and radius constraints \cite{Alford:2006vz,Klahn:2006iw,Blaschke:2007ri,Grunfeld:2007jt}. This offers still a broad spectrum of possible realizations of hybrid stars in nature, as classified recently in Ref.~\cite{Alford:2015dpa}. The two extreme scenarios are the {\em masquerade} case \cite{Alford:2004pf}, where the corresponding quark-hadron hybrid stars appear to have almost identical static properties to pure neutron stars, and the {\em high-mass twin} case \cite{Blaschke:2013ana,Benic:2014jia} associated with a strong first-order phase transition. 
The latter can be identified by observing CSs with similar high masses (such as PSR J1614-2230 
\cite{Demorest:2010bx,Fonseca:2016tux} and PSR J0348+0432 \cite{Antoniadis:2013}) but significantly different radii. This requires an accuracy of radius measuments of about 500 m,  as it shall be provided by the NICER mission of NASA \cite{nicer}, planned for launch in the near future. 
In the case of a smooth cross-over transition, i.e. the masquerade case, precise observations of CS mass and/or radii will not allow to provide evidence for the existence of quark matter at their interiors. 
In such a situation, the transport properties of dense matter may provide the decisive diagnostic tool via the cooling history of CS.

Besides cold CSs, also in protoneutron stars (PNS) the transport properties play a crucial role. PNSs are born hot and lepton rich in the violent event of a core-collapse supernova. They deleptonize and cool on a timescale on the order of 10--30~s via the emission of neutrinos of all flavors \cite{Pons:1998mm,Fischer:2010,Huedepohl:2010, Roberts:2012, MartinezPinedo:2012rb, Fischer:2016}. The appearance and role of quark matter in PNS and core-collapse supernovae has long been studied by means of conducting numerical studies \cite{Pons:2001,Nakazato:2008su,Sagert:2008ka,Fischer:2011}, also in particular as the trigger of the actual supernova explosion via a strong first-order phase transition at high density. This launches a strong hydrodynamics shock wave, in addition of the standard supernova standing bounce shock, and the release of an outburst of neutrinos of all flavors \cite{Dasgupta:2009yj}. Those neutrinos are released during the shock passage across the neutrinospheres of last scattering, located always at low densities where matter is composed of hadronic degrees of freedom. The future observation of such neutrino signal may reveal yet unknown details of the associated with the quark-hadron phase transition. The caveat in all these studies was the treatment of neutrino interactions in quark matter, which was treated at the level of nucleons only. This approximation is valid when temperatures are on the order of 10~MeV or above. However, during the long-term evolution of deleptonizing protoneutron star, as the core temperature decreases below about 1~MeV, weak interactions at the quark level become important. Unlike in studies of cooling CS, where neutrino-quark interactions are treated at different levels of sophistication \cite{Iwamoto:1980eb,Iwamoto:1982,Haensel:1986}, for supernova studies the general framework has to be derived 
along the lines of Refs.~\cite{Burrows:1980,Iwamoto:1983}.

Since the cooling and spin evolution of CS depends sensitively on the thermal and transport properties of dense matter, the latter can be determined from the observation of cooling CSs, with particular emphasis on young objects like Cassiopeia~A \cite{Ho:2009mm}. For a recent discussion of the role of the stiffness of the EoS and the superfluidity gaps in this context, cf. Refs.~\cite{Ho:2014pta,Grigorian:2016leu}. If quark matter is present in the CS interior we expect it to be in a color-superconducting state which entails a strong dependence on the pairing pattern and the sizes of pairing gaps. In the present study, we will focus on the discussion of direct Urca neutrino emissivities and bulk viscosities of color-superconducting quark matter. The numerical analysis is based on a Nambu-Jona-Lasinio (NJL) type model,  allowing a consistent description of the density and temperature dependent quark masses, pairing gaps and chemical potentials under neutron star constraints. The resulting phase diagram suggests that three-flavor phases of the color-flavor-locking (CFL) type occur only at rather high densities \cite{Ruester:2005jc,Blaschke:2005uj} and render hybrid star configurations gravitationally unstable \cite{Buballa:2003qv,Klahn:2006iw}. Moreover, due to large pairing gaps in CFL quark matter, the r-mode  instabilities cannot be damped  \cite{Madsen:1999ci} and cooling is inhibited \cite{Blaschke:1999qx}.

By this reasoning, we shall focus on two-flavor quark matter as the relevant case for discussion of quark deconfinement in CSs as well as in the protoneutron star evolution during supernova collapse.
Due to the pairing instability the scalar antitriplet diquark correlations form a condensate in the color
superconducting 2SC phase with a critical temperature $T_{\rm 2SC}$ that is on the order of $20 - 50$ MeV \cite{Ruester:2005jc,Blaschke:2005uj}.  
Within the Polyakov-loop extension of the NJL model, this temperature may even reach up to the
pseudocritical temperature $T_c=154$ MeV found in recent lattics QCD simulations 
 \cite{Borsanyi:2013bia,Bazavov:2014pvz} for the chiral and
Polyakov-loop transition at vanishing baryon number densities, see 
\cite{Blaschke:2010ka,Ayriyan:2016lbx}.

The standard 2SC phase, however, pairs only two of the three colors (e.g., red-green) of quarks, leaving color unpaired (blue quarks in this example) on which then the rapid direct Urca cooling process may proceed, too rapid in comparison with compact star phenomenology.
This problem has prompted the introduction of a purely phenomenological gap 
(X-gap) for the quarks of the unpaired color \cite{Grigorian:2004jq}.
For a recent investigation of such a fully gapped 2SC phase see 
\cite{Sedrakian:2015qxa,Sedrakian:2013xgk}, which may be contrasted to the transport 
\cite{Alford:2014doa} and cooling properties  \cite{Hess:2011qw} in the original 2SC phase.
In this context also the anisotropic crystalline color superconductivity phases have been discussed,
which have been reviewed in  \cite{Alford:2007xm,Anglani:2013gfu}.

It is an unsatisfactory situation to have no candidate for the microscopic pairing pattern that could justify the phenomenological X-phase in the 2SC+X model of the fully gapped 2SC phase.
One alternative is provided by the isotropic color-spin-locking (iso-CSL) phase suggested in 
\cite{Aguilera:2005tg,Aguilera:2006cj} 
modifying earlier work on spin-1 color superconducting phases
\cite{Schmitt:2004et}.
The iso-CSL phase is a single flavor pairing scheme and therefore rather inert against isospin asymmetry and strong magnetic fields, thus qualifying as a robust pairing pattern for compact star applications.
Technically the description of the transport and cooling properties of this phase follows that of the family
of spin-1 color superconductors which have been studied in detail in \cite{Schmitt:2005wg}.

In the present work, we will focus on two-flavor color-superconducting phases in CSs, the 2SC+X phase of Ref.~\cite{Grigorian:2004jq}, for which a detailed investigation of the cooling phenomenology for hybrid stars has already been worked out \cite{Popov:2005xa,Blaschke:2006gd}, and the iso-CSL phase \cite{Aguilera:2005tg,Aguilera:2006cj} for which a consistent microscopic calculation of the direct Urca emissivity and the bulk viscosity will be presented here for the first time \cite{Blaschke:2007bv}. This will form the basis of further phenomenological in astrophysics, with applications to supernovae and CSs. 


\section{Thermodynamics of iso-CSL and 2SC phases}
 
One can introduce a general thermodynamical potential which is in 
mean field approximation  
\begin{eqnarray}\label{GKTP}
&~&\Omega(\mu_B,\mu_Q,\mu_8,T)=\frac{\bar{\sigma}_u^2+\bar{\sigma}_d^2}{8G_S}+
\frac{\Delta_u^2+\Delta_d^2}{8G_D}  \\
~&&-2 \int \frac{d^3~p}{(2\pi)^3}\sum
\limits_{i=1}^{12}\biggl[\frac{\lambda_{i}}{2}+T{\rm ln}(1+e^{-\lambda_{i}/T})\biggr]+\Omega_l-\Omega_0, \nonumber
\end{eqnarray}
where $\lambda_{i}$ are the excitation energies for the corresponding modes.\\
Here $\Omega_e=-\mu_Q^4/12\pi^2-\mu_Q^2T^2/6-7\pi^2T^4/180$ denotes the 
thermodynamic potential of ultra-relativistic electrons, where $\mu_Q=-\mu_e$, 
and  $\Omega_0$ is the divergent vacuum contribution which has to be 
subtracted to assure vanishing energy and pressure of the vacuum.

\subsection{iso-CSL phase}

The gap matrix of the iso-CSL phase \cite{Aguilera:2005tg,Aguilera:2006cj}  is 
\begin{equation}
  \hat{\Delta}=\Delta(\gamma_3\lambda_2+\gamma_2\lambda_5+\gamma_1\lambda_7),
\end{equation}
a scalar product of the three antisymmetric color matrices with the spin
matrices ($\gamma_3,\gamma_2,\gamma_1$) \cite{Aguilera:2005tg}, 
whereas the pairing pattern for the 2SC phase is 
\begin{equation}
  \hat{\Delta}=\Delta(i\gamma_5\tau_2\lambda_2), 
\end{equation}
coupling two different flavor with each other.  
\\
Note that in the iso-CSL phase all modes have a gap in the corresponding
 excitation spectra 
\begin{equation}\label{En1}
\lambda_{1}^2=(\epsilon_{u,eff}(p)-\mu_{u,eff}(p))^2+\Delta_{u,eff}^2(p),
\end{equation}
with the effective values
\begin{align}
\epsilon_{u,eff} &= \sqrt{p^2+M_{u,eff}^2(p)},\nonumber\\
M_{u,eff}(p) &=\frac{\mu_u}{\mu_{u,eff}(p)}M_u(p), \nonumber\\
\mu_{u,eff}(p) &= \mu_u\sqrt{1+\Delta_u^2/(\mu_u)^2},\nonumber\\
\Delta_{u,eff}^2(p)&=a_{u,1}\Delta_u^2.
\end{align}
and 
\begin{equation}
\lambda_{3,5}^2(p)=(\epsilon_u(p)-\mu_u)^2+a_{u,(3,5)}(p)\Delta_u^2,
\end{equation}
with the momentum-dependent coefficients 
\begin{eqnarray}\label{En2}
a_{u,1}(p)&=&\frac{M_u^2(p)}{\mu_{u,eff}^2(p)}\nonumber\\
a_{u,(3,5)}(p)&=&\frac{1}{2}\Biggl[5-\frac{p^2}{\epsilon_u(p)\mu_u} \\
&&\pm \sqrt{\left(1-\frac{p^2}{\epsilon(p)\mu_u}\right)^2+\frac{8~M_u^2(p)}{
\epsilon_u^2(p)}}~\Biggl],\nonumber
\end{eqnarray}
where $\epsilon_u(p)=\sqrt{p^2+M_u^2(p)}$.\\
The excitation energies $\lambda_{7-12}$ are obtained by changing the flavor
 ($u\rightarrow d$) and the even modes by exchange of  
$\mu \rightarrow -\mu$ in Eqs.~(\ref{En1}-\ref{En2}).

\subsection{2SC phase}
In case of the 2SC phase four out of the twelve eigenvalues $\lambda_a$ belong
to the ungapped blue quarks and are determined easily via textbook methods 
\cite{Kapusta:2006pm} as $\lambda_{1..4} = \epsilon_f(p)\pm \mu_{fb}$. 
Here the dispersion relation $\epsilon_f(p)=\sqrt{p^2+M_f^2(p)}$ contains the 
dynamical mass function $M_f(p)=m_f+\phi_f$ for the two quark flavors $f=u,d$.\\
We have introduced the chemical potentials for the quarks of unpaired color
$\mu_{ub}= \mu_B/3+2\mu_Q/3-2\mu_8/2$ and $\mu_{db}=\mu_{ub}-\mu_Q$.\\  
The other eight eigenvalues $\lambda_{5-12}$ belong to the red and green
 quarks which are paired in the 2SC state and have therefore an identical
 eigenvalue spectrum.
It is thus sufficient to determine the four eigenvalues for the real and symmetric matrices
of the red quarks 
\begin{equation}\label{eq50}  
\mathcal{M}_s=\begin{vmatrix}  -\mu_{d,r}+M_d & p & 0 & -\Delta  \\ 
 p & -\mu_{d,r}-M_d & \Delta & 0 \\ 0 & \Delta & \mu_{u,r}+M_u &
 p \\ -\Delta & 0 & p & \mu_{u,r}-M_u \end{vmatrix}  \\
\end{equation}  
with $s=\pm$  (similar to the one discussed in 
\cite{Blaschke:2005uj,Ruester:2006yh}) for the CFL phase).
The eigenvalues of the matrix (\ref{eq50}) can be found as the roots of the
characteristic polynomial  
\begin{equation}  
\lambda^4+a_3\lambda^3+a_2\lambda^2+a_1\lambda+a_0=0  
\end{equation}   
with the coefficients   
\begin{align}  
a_0 &=\Delta^4+2\Delta^2(M_dM_u+\mu_{d,r}\mu_{u,r}+p^2)\nonumber\\
&~~~~~~+[M_d^2-(\mu_{d,r})^2+p^2][M_u^2-(\mu_{u,r})^2+p^2]\nonumber\\  
a_1 &=-2(M_u^2\mu_{d,r}+\Delta^2(\mu_{d,r}-\mu_{u,r})-M_d^2\mu_{u,r}\nonumber\\
&~~~~~~+(\mu_{d,r}-\mu_{u,r})(\mu_{d,r}\mu_{u,r}+p^2))\nonumber\\  
a_2 &=-2\Delta^2-M_d^2-M_u^2+(\mu_{d,r})^2-4\mu_{d,r}\mu_{u,r}\nonumber\\
&~~~~~~+(\mu_{u,r})^2-2p^2\nonumber\\  
a_3 &=2(\mu_{d,r}-\mu_{u,r}).  
\end{align}    
The four roots of the quartic equation can be calculated by use of the real
 solution of the cubic equation \cite{Abramowitz:1984}    
\begin{equation}  
u^3-a_2u^2+(a_1a_3-4a_0)u-(a_1^2+a_0a_3^2-4a_0a_2)=0  ~.
\end{equation}  
The real roots of the quartic equation are then the root of the quadratic
 equation   
\begin{equation}  
v^2+\left[\frac{a_3}{2}\mp\left(\frac{a_3^2}{4}+u_1-a_2\right)^{\frac{1}{2}}
\right]v+\frac{u_1}{2}\mp\left[\left(\frac{u_1}{2}\right)^2-a_0\right]^{
\frac{1}{2}}=0.  
\end{equation}  

\subsection{Comparison}
The global minima of $\Omega(\mu_B,\mu_Q,\mu_8,T)$ in the space of the order
parameters corresponds to the thermodynamical equilibrium and is solution
of the following gap-equations
\begin{equation}
\frac{\partial{\Omega(T,\mu_f)}}{\partial{\bar{\sigma}_f}}=\frac{\partial{
\Omega(T,\mu_f)}}{\partial{\Delta}_f}=\frac{\partial{\Omega(T,\mu)}}{
\partial{\bar{\omega}}}=\frac{\partial{\Omega(T,\rho)}}{\partial{\bar{\rho}}}
=0.
\end{equation}
We investigate the phase diagram for the isotropic iso-CSL 
and for the 2SC phase by use of
the M(p=0)=380 MeV parameter set of \cite{Grigorian:2006qe}. \\ 
The order parameter and the chemical potential of 
up-,down-quarks and electrons are shown (Figure \ref{CSL1}) in dependence of
the quark chemical potential $\mu$ for temperature $T=0$.
If the 2SC-phase is partial suppressed or breaks up completely, then the
iso-CSL phase is realised and the phase diagram for quark matter has the form
 of Figure \ref{CSL2}.

\begin{widetext}

\begin{figure*}[htp!] 
\includegraphics[width=0.475\columnwidth]{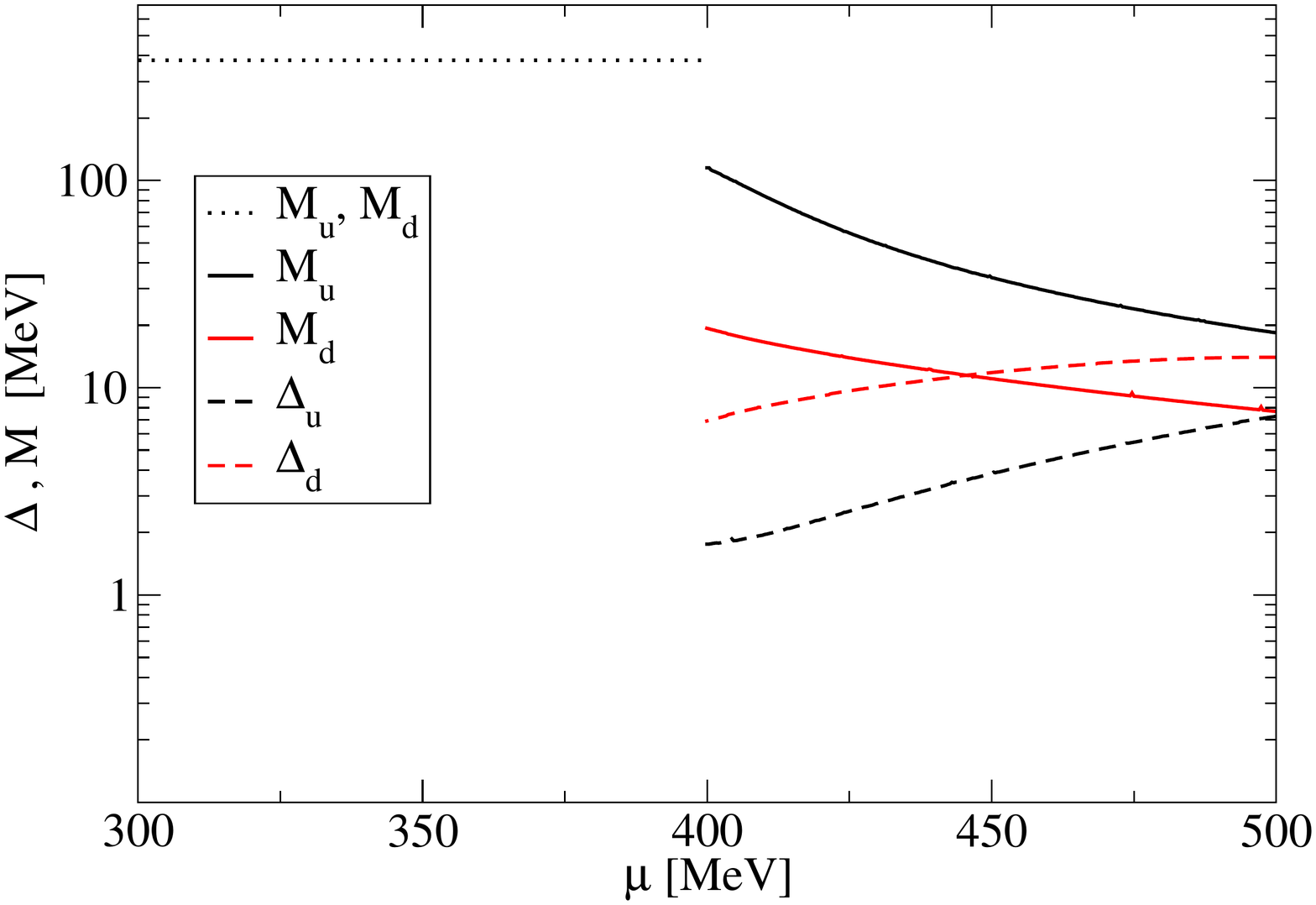}
\hfill
\includegraphics[width=0.475\columnwidth]{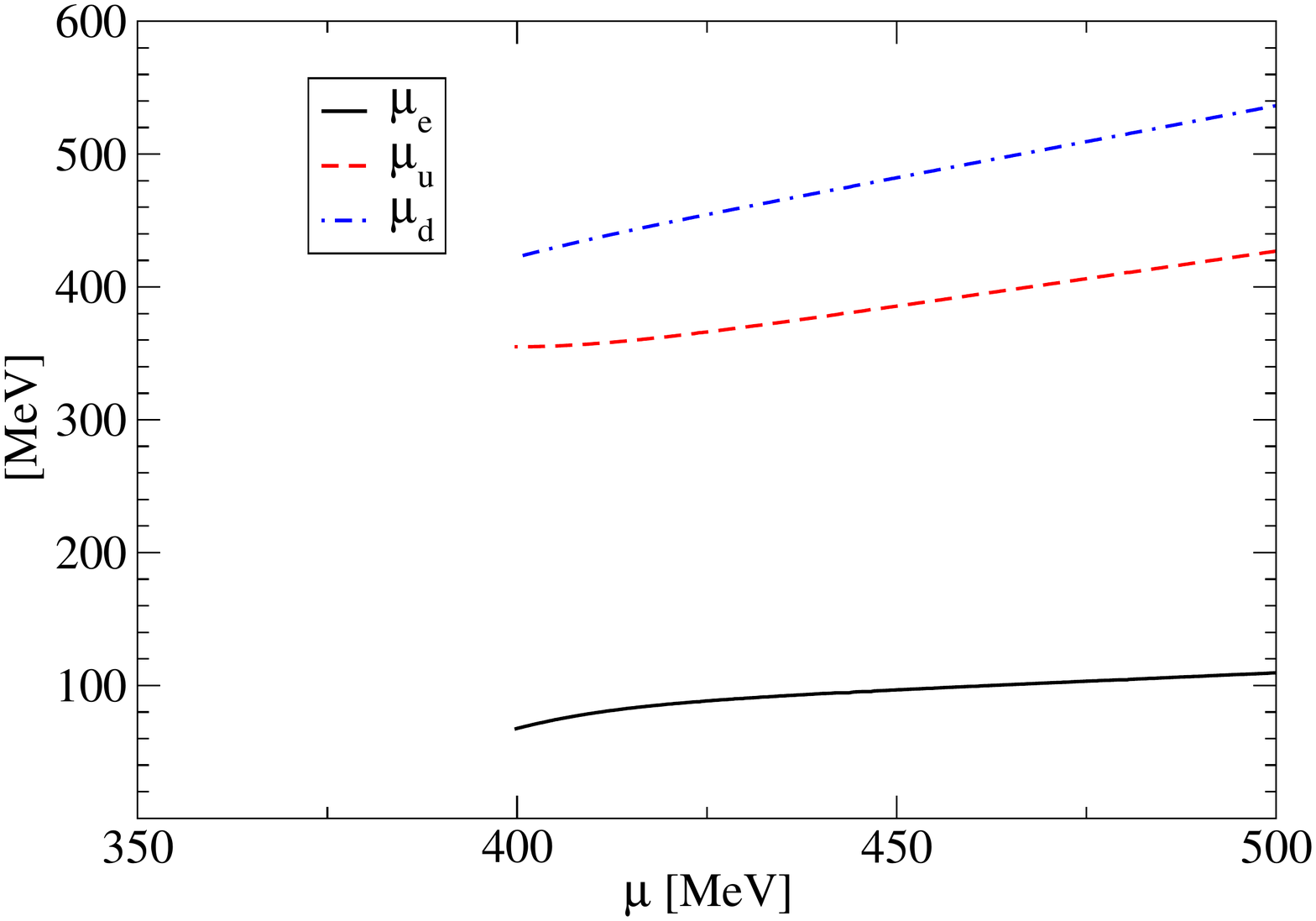}
\caption{The left panel shows the dynamical quark masses and pairing gaps ($\eta_D=3/8$) 
for the iso-CSL phase as function of the quark-chemical potential and
the corresponding chemical potentials for quarks and electrons are given in the right panel. 
\label{CSL1}}
\end{figure*}

\begin{figure*}[htp!]
\includegraphics[width=0.475\columnwidth]{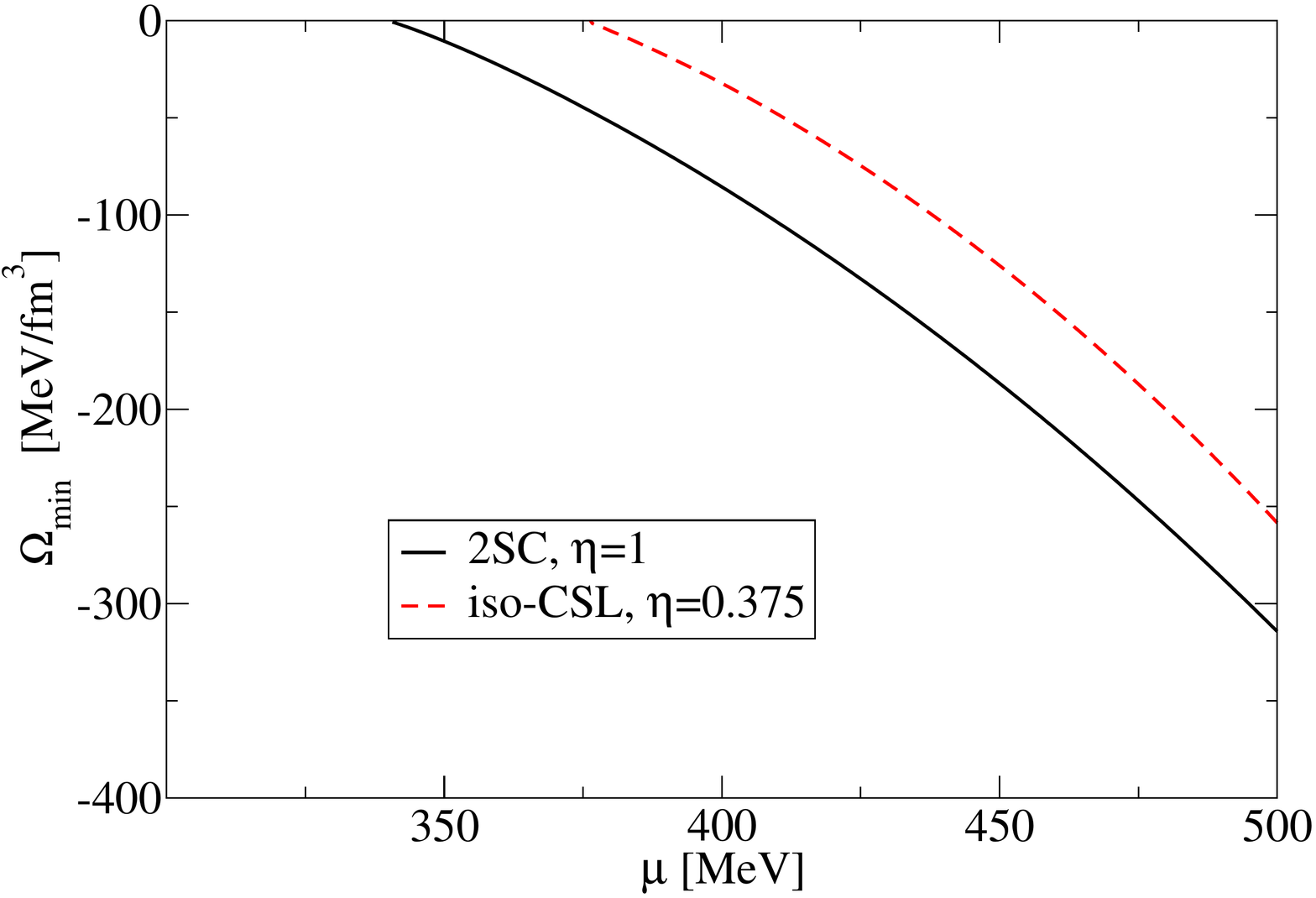}
\hfill
\includegraphics[width=0.475\columnwidth]{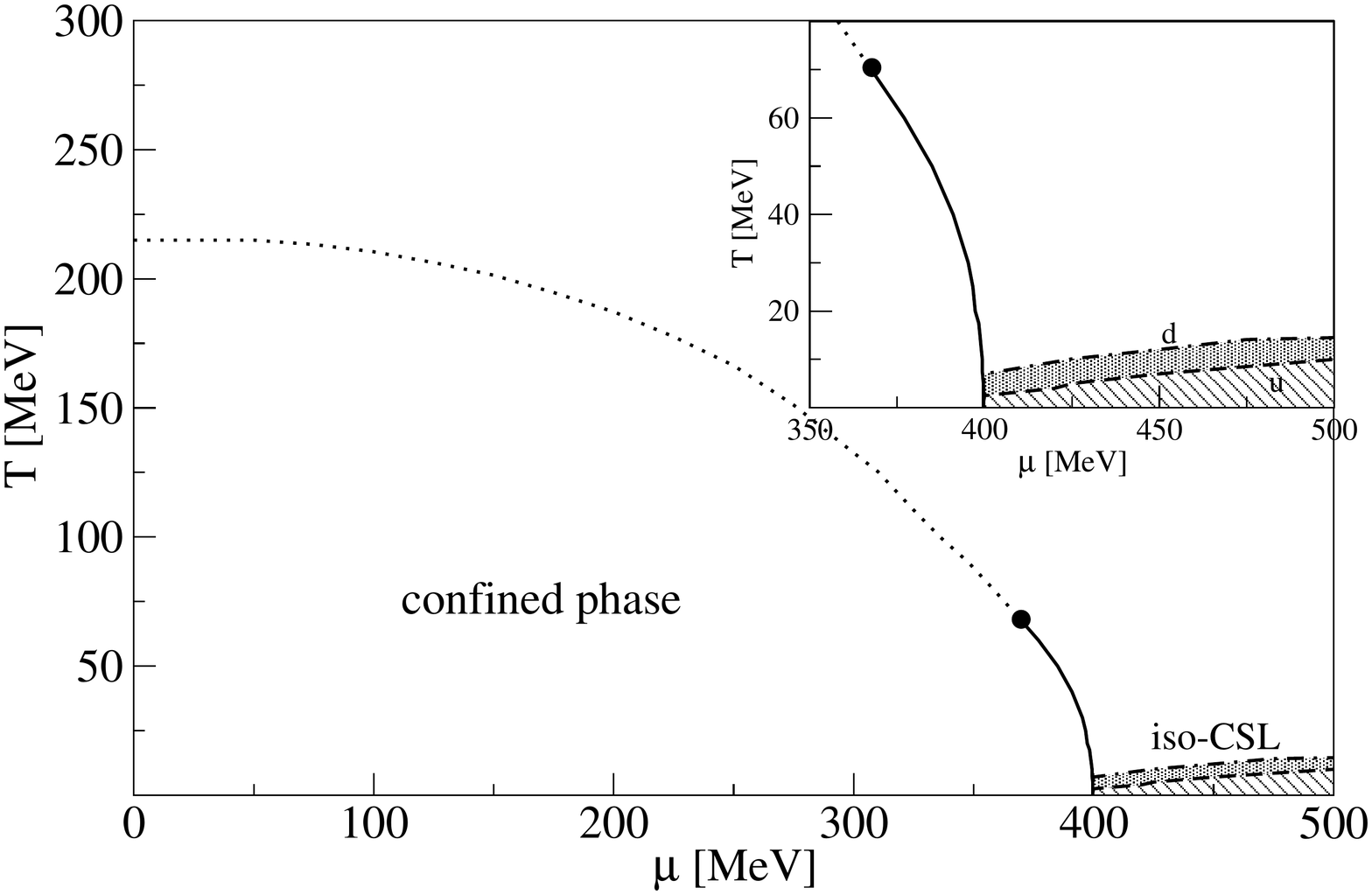}
\caption{
The left panel displays the minima of the thermodynamical potential for 2SC
and iso-CSL quark matter at $T=0$ as a function of the quark chemical potential.
In the right panel the iso-CSL Phase diagram calculated with the NJL form factor 
and a diquark coupling of $\eta_D=3/8$ is shown.
\label{CSL2}}
\end{figure*}


\section{Transport properties}

For the derivation of the kinetic equation for neutrinos in the iso-CSL and 2SC+X phases of color superconducting quark matter and the corresponding direct Urca neutrino emissivities we follow 
the steps outlined in \cite{Schmitt:2005wg,Jaikumar:2005hy,Wang:2006tg}, while for the discussion 
of the bulk viscosity of these phases we adopt the formulation given in Ref.~\cite{Sa'd:2006qv}.

\subsection{Kinetic equation for neutrinos in warm dense quark matter}
The kinetic equation for neutrino transport in the Green's function representation,
\begin{equation} \label{1.1}
i\partial_x^{\alpha}{\rm Tr_D}[\gamma_{\alpha}G_{\nu}^<(X,q_2)]=
-{\rm Tr}[G_{\nu}^>(X,q_2)\Sigma^<_{\nu}(X,q_2)-\Sigma^>_{\nu}(X,q_2)
G_{\nu}^<(X,q_2)]~,
\end{equation}
can be derived from the Kadanoff-Baym formalism \cite{Kadanoff:1962} by a 
gradient expansion, which is valid if the neutrino Green's functions  
\begin{eqnarray} \label{1.2}
&&iG_{\nu}^<(t,q_2)=-(\gamma^{\beta}q_{2,\beta}+\mu_{\nu}\gamma_0)
\frac{\pi}{q_2}\{f_{\nu}(t,{\bf q_2})\delta(p_2^0+\mu_{\nu}-|{\bf q_2}|)
-[1-f_{\bar{\nu}}(t,-{\bf q_2})]\delta(q_2^0+\mu_{\nu}+|{\bf q_2}|)\}
\nonumber\\
&&iG_{\nu}^>(t,q_2)=~~(\gamma^{\beta}q_{2,\beta}+\mu_{\nu}\gamma_0)
\frac{\pi}{q_2}\{[1-f_{\nu}(t,{\bf q_2})]\delta(q_2^0+\mu_{\nu}-|{\bf q_2}|)-
f_{\bar{\nu}}(t,-{\bf q_2})\delta(q_2^0+\mu_{\nu}+|{\bf q_2}|)\}
\nonumber\\
\end{eqnarray}
and the neutrino self energies
\begin{eqnarray} \label{1.3}
&&\Sigma_{\nu}^<(t,q_2)=\frac{G_F^2}{2}\int\frac{d^4~q_1}{(2\pi^4)}
\gamma^{\mu}(1-\gamma_5)(\gamma^{\alpha}q_{1,\alpha}+\mu_e\gamma_0)
\gamma^{\nu}(1-\gamma_5)\Pi_{\mu\nu}^{>}(q_1-q_2)\frac{\pi}{q_1}
f_e(t,{\bf q_1})\delta(q_1^0+\mu_e-|{\bf q}_1|),
 \nonumber\\
&~&\Sigma_{\nu}^>(t,q_2)=\frac{G_F^2}{2}\int\frac{d^4~q_1}{(2\pi^4)}
\gamma^{\mu}(1-\gamma_5)(\gamma^{\alpha}q_{1,\alpha}+\mu_e\gamma_0)
\gamma^{\nu}(1-\gamma_5)\Pi_{\mu\nu}^{<}(q_1-q_2)\frac{\pi}{q_1}
[1-f_e(t,{\bf q_1})]\delta(q_1^0+\mu_e-|{\bf q}_1|)
\nonumber\\
\end{eqnarray}
are slowly varying functions of the space-time coordinate $X=(t,{\bf x})$.
The functions $\Pi^{<,>}_{\mu\nu}(q_1-q_2)$ are the self-energies of the W-bosons.
The W-boson exchange can be expressed in its local form due to the smallness of the  
neutrino-energy compared to the W-boson mass. 
We follow the steps of \cite{Schmitt:2005wg} to obtain the time dependent neutrino distribution function
\begin{equation}
\label{6.2}
\frac{\partial}{\partial t}f_{\nu}(t,{\bf q_2})=\frac{G^2_F}{8}\int\frac{d^3 
{\bf q_1}}{(2\pi)^3~p_{F,e}~p_{F,\nu}}\mathcal{L}^{\mu\nu}(q_1,q_2)
~n_F(p_{F,e}-\mu_e)n_B(p_{F,\nu}+\mu_e-p_{F,e}){\rm Im}\Pi_{\mu\nu}^R(q)
\end{equation}
from Eq.~(\ref{1.1}), where 
\begin{equation}
\label{6.2a}
\mathcal{L}^{\mu\nu}(q_1,q_2)\equiv {\rm Tr} 
\left[(\gamma_0q_1^0-{\vec{\gamma}}\cdot {\bf q_1})\gamma^\mu(1-\gamma^5) 
(\gamma_0q_2^0-{\vec{\gamma}}\cdot {\bf q_2})\gamma^\nu(1-\gamma^5) 
\right]
\end{equation}
is the leptonic tensor.  
We insert the imaginary part of the polarization tensor Eq.~(\ref{4.10}) and obtain
\begin{eqnarray}
\label{6.3a}
\frac{\partial}{\partial t}f_{\nu}(t,{\bf q_2})&=&-\frac{G^2_F~\pi}{8}\cos^2 
\theta_c \int\frac{d^3 {\bf q_1}}{(2\pi)^3~p_{F,e}~p_{F,\nu}}\int\frac{d^3 
{\bf p}}{(2\pi)^3}~n_F(p_{F,e}-\mu_e)n_B(p_{F,\nu}+\mu_e-p_{F,e})\nonumber\\
&\times&\sum\limits_{r=1,3,5}\left[2A_r^{\ast}(E_p,E_k)\mathcal{L}^{\mu\nu}
(q_1,q_2)~\mathcal{H}_{\mu\nu}^{(n)}(\hat{p},\hat{k})-\Delta^2~B_r^{\ast}
(E_p,E_k)\mathcal{L}^{\mu\nu}(q_1,q_2)~\mathcal{H}_{\mu\nu}^{(a)}(\hat{p},\hat{k})\right].
\end{eqnarray}
To simplify this expression we can neglect the anomalous contribution
 $\mathcal{H}_{\mu\nu}^{a}$, which is small compared to the normal part 
$\mathcal{H}_{\mu\nu}^{n}$ \cite{Jaikumar:2005hy} and obtain
\begin{eqnarray}
\label{6.3b}
\frac{\partial}{\partial t}f_{\nu}(t,{\bf q_2})&=&-\frac{G^2_F~\pi}{4}\cos^2
 \theta_c \int\frac{d^3 {\bf q_1}}{(2\pi)^3~p_{F,e}~p_{F,\nu}}\int\frac{d^3
 {\bf p}}{(2\pi)^3}~n_F(p_{F,e}-\mu_e)\nonumber\\
&\times&\sum\limits_{r=1,3,5}\sum\limits_{e_1,e_2=\pm}B_p^{e_1}B_k^{e_2}
n_F(-e_1\xi_{p,r})n_F(e_2\xi_{k,r})\delta(q_0+e_1\xi_{p,r}-e_2\xi_{k,r})
 \mathcal{L}^{\mu\nu}(q_1,q_2)~\mathcal{H}_{\mu\nu}^{(n)}(\hat{p},\hat{k}),
\nonumber\\
\end{eqnarray}
where the Bose functions cancel each other, because their argument represents
the momentum transfer of the W-Boson 
($p_{F,\nu}+\mu_e-p_{F,e}=-q_0=\xi_{p,r}-\xi_{k,r}$).
Contraction between leptonic and hadronic tensor leads to 
\begin{eqnarray}
\label{Ltensor}
\mathcal{L}^{\mu\nu}(q_1,q_2)\mathcal{H}_{\mu\nu}^{(\rm n)}(\hat{p},\hat{k}) 
&=& 64 q_1^0 q_2^0 (1-\hat{q}_1\cdot\hat{p})(1-\hat{q}_2\cdot\hat{k})
=64 q_1^0 q_2^0 (1-\cos\theta_{ue})(1-\cos\theta_{\nu d}).
\end{eqnarray} 
In case of small angles 
$1-\cos\theta_{ue} 
\simeq  \theta^2_{ue}$ 
and one obtains the simple expression 
\begin{equation}
\mathcal{L}^{\mu\nu}(q_1,q_2)\mathcal{H}_{\mu\nu}^{(\rm n)}(\hat{p},\hat{k}) 
= 64 q_1^0 q_2^0 \theta^2_{ue}(1-\cos\theta_{\nu d}) = 64~p_{F,e}~p_{F,\nu}
~\theta^2_{ue}(1-\cos\theta_{\nu d}).
\end{equation} 
The angle $\theta^2_{ue}$ can be expressed by the angle $\theta^2_{de}$, see Appendix~\ref{D},   
Eq.~(\ref{5.7}) for the perturbative and the quark mass effect. 
 
We concentrate on the perturbative treatment, where 
$\theta_{de}^2=(4/3\pi)\alpha_s$ and the Boltzmann equation for the time evolution of the neutrino
distribution function becomes 
\begin{eqnarray}\label{neutrino3}
\frac{\partial}{\partial t}f_{\nu}(t,{\bf q_2})&=&-\frac{64}{3}~\alpha_s~G^2_F
~\cos^2 \theta_c \int\frac{d^3 {\bf q_1}}{(2\pi)^3}\int\frac{d^3 {\bf p}}{
(2\pi)^3}~n_F(p_{F,e}-\mu_e) \sum\limits_{r=1,3,5}\sum\limits_{e_1,e_2=\pm}
B_p^{e_1}B_k^{e_2}\nonumber\\
&~&\times n_F(-e_1\xi_{p,r})n_F(e_2\xi_{k,r})\delta(p_{F,e}-\mu_e-p_{F,\nu}+
e_1\xi_{p,r}-e_2\xi_{k,r})(1-\cos\theta_{\nu d}).
\end{eqnarray}
Here the replacement $q_0=p_{F,e}-\mu_e-p_{F,\nu}$ has been made in the $\delta$-function 
of Eq.~(\ref{6.3b}).
The $\delta$-function of Eq.~(\ref{neutrino3}) vanishes if the angle between
 up- and down-quarks $\theta_{ud}$ corresponds to a fixed value $\theta_0$.
 The value of the angle $\theta_0$ ist given by $\cos\theta_0\equiv 1-\kappa
\mu_e^2/(\mu_u\mu_d)$, with $\kappa\equiv 2\alpha_s/(3\pi)$.
 The angle $\theta_0$ is independent of the neutrino Fermi momentum $p_{F,\nu}$. 
The $\delta$-function can be replaced now by
 $\mu_e/(\mu_u\mu_d)\delta(\cos\theta_{ud}-\cos\theta_0)$ 
\cite{Schmitt:2005wg}.
Eq.~(\ref{neutrino3}) becomes 
\begin{eqnarray}\label{neutrino4}
\frac{\partial}{\partial t}f_{\nu}(t,{\bf q_2})&\simeq&-\frac{64}{3}~\alpha_s~
G^2_F~ \cos^2 \theta_c~ \mu_e\mu_u\mu_d\int\frac{dkd\Omega_k}{(2\pi)^3}\int
\frac{dpd\Omega_p}{(2\pi)^3}(1-\cos\theta_{\nu d})\delta(\cos\theta_{ud}-
\cos\theta_0)\nonumber\\
&~&~~\times \sum\limits_{r=1,3,5}\sum\limits_{e_1,e_2=\pm}B_p^{e_1}B_k^{e_2}
 n_F(-e_1\xi_{p,r})n_F(e_2\xi_{k,r})n_F(p_{F,\nu}-\xi_{p,r}^-+\xi_{k,r}^-),
\end{eqnarray}
where the variable of the integration  ${\bf q_1}$ are changed to
 ${\bf k}\equiv {\bf p} + {\bf q}_1 - {\bf q}_2$ and the phase space element
 for masseless quarks can be written as $d^3{\bf p}=\mu_u^2 dpd\Omega_p$.
The argument in the Fermi-distribution is replaced, because 
 $p_{F,e}-\mu_e=p_{F,\nu}-\xi_{p,r}^-+\xi_{k,r}^-$. 
Introducing dimensionless variables  
\begin{equation}
x\equiv \frac{k-\mu_d}{T},~~~~~y\equiv\frac{p-\mu_u}{T},
~~~~~z\equiv\frac{q_2}{T}. 
\end{equation}
changes the integration range $-\mu_{u,d}/T$ to $\infty$.
As long as the main contribution results from  $x,y\ll\mu_{u,d}/T$ the result
 of the integration is not affected by a shift of the lower boundary to
 $-\infty$.
 Therefore the uneven parts in the Bogoliubov coefficients 
\begin{eqnarray}
B_{{\bf k},r,d}^{e_2}&=&\frac{1}{2}-\frac{e_2x}{2\sqrt{x^2+\lambda_{{\bf k},r}
\Delta_d^2}}~,\nonumber\\
B_{{\bf p},r,u}^{e_2}&=&\frac{1}{2}-\frac{e_1y}{2\sqrt{y^2+\lambda_{{\bf p},r}
\Delta_u^2}}~,
\end{eqnarray}
can be neglected in the integrand, thus only the even parts, a constant term
 1/2 remains \cite{Schmitt:2005wg}.  
The integral over x and y can now be done in the range from 0 to $\infty$,
 where one gets a factor 2 from the interval $-\infty \rightarrow 0$, which 
cancels with the constant term 1/2 from the Bogoliubov coefficients.   
Therefore, Eq.~(\ref{neutrino4}) becomes
\begin{equation}\label{neutrino5}
\frac{\partial}{\partial t}f_{\nu}(t,{\bf q_2})\simeq-\frac{64}{3}~\alpha_s~
G^2_F~ \cos^2 \theta_c~ \mu_e\mu_u\mu_d~T^2 \sum\limits_{r=1,3,5}
 \int\frac{d\Omega_k}{(2\pi)^3}\int\frac{d\Omega_p}{(2\pi)^3}(1-\cos\theta_{
\nu d})\delta(\cos\theta_{ud}-\cos\theta_0)\mathcal{F}_r(z),
\end{equation}
with
\begin{eqnarray}
\label{integral1}  
\mathcal{F}_r(z) = \sum\limits_{e_1,e_2=\pm}\int_0^{\infty}\int_0^{\infty}  
dxdy\ (e^{-e_1\sqrt{y^2+a_{u,r}\Delta_u^2}}+1)^{-1}(e^{e_2\sqrt{x^2+a_{d,r}
\Delta_d^2}}+1)^{-1}
(e^{z+e_1\sqrt{y^2+a_{u,r}\Delta_u^2}-e_2\sqrt{x^2+a_{d,r}\Delta_d^2}}+1)^{-1}.
\nonumber\\
\end{eqnarray}

\subsection{Emissivity}  
The loss of energy by neutrinos per unit of time and volume is given by
\begin{equation} 
\label{Neutrinoemiss}
\varepsilon_{\nu}\equiv -\frac{\partial}{\partial{t}}\int\frac{{\rm d}^3{\bf 
q_2}}{(2\pi)^3}|{\bf q_2}|[f_{\nu}(t,{\bf q_2})+f_{\bar{\nu}}(t,{\bf q_2})]
= -2\frac{\partial}{\partial{t}}\int\frac{{\rm d}^3{\bf q_2}}{(2\pi)^3}~
p_{F,\nu}~f_{\nu}(t,{\bf q_2}).
\end{equation}
The corresponding expression for the neutrino emissivity can be obtained by
 inserting Eq.~(\ref{neutrino5}) into Eq.~(\ref{Neutrinoemiss}). 
Performing the angle integration gives a factor $32\pi^3$ 
\cite{Iwamoto:1980eb}, summation over the color states $r$ 
a factor 3 and by use of the integral 
\begin{equation}
\sum\limits_{e_1,e_2=\pm}\int_0^{\infty}dzz^3\int_0^{\infty}dx\int_0^{\infty}
dy(e^{-e_1y}+1)^{-1}(e^{e_2x}+1)^{-1}(e^{z+e_1y-e_2x}+1)^{-1}=\frac{457}{5040}
\pi^6
\end{equation}
\end{widetext}
one obtains the gapless result of Iwamotos seminal paper \cite{Iwamoto:1982}
 where the direct Urca emissivity of quark matter,  
\begin{equation} 
\epsilon_0=\frac{457}{630}\alpha_s G_F^2~\mu_e\mu_u\mu_d~ T^6~,  
\end{equation} 
was derived for the first time.
Since then, there have been a number of calculations, in particular for color superconducting phases, 
we refer to \cite{Jaikumar:2005hy,Schmitt:2005wg,Wang:2006tg}. 
However, none of these is useable for cooling simulations because they  
have either ungapped modes which result in too fast cooling or the pairing  
pattern is not microscopically founded.  
Nevertheless, in deriving the neutrino emissivities for the 2SC+X and the  
iso-CSL phase we follow the strategy of these References by using the form 
\begin{eqnarray}\label{emissi}  
\epsilon_{\rm Urca}&=& \epsilon_0~ G_3(\Delta_u,\Delta_d),  
\end{eqnarray}  
where we introduced the function     
\begin{eqnarray}  
G_n(\Delta_u,\Delta_d)&=& 
\frac{5040}{1371\pi^6}\int\limits_0^{\infty}dz~z^n  
\left[\mathcal{F}_1(z)+\mathcal{F}_3(z)+\mathcal{F}_5(z)\right] ~, 
\nonumber \\
\end{eqnarray}  
with Eq.~(\ref{integral1}) characterising the influence of the
 superconducting gaps on the corresponding emissivity.  
For the iso-CSL phase, the coefficients $a_{u,r}$ and $a_{d,r}$ for $r=1,3,5$  
are defined in Ref.~\cite{Aguilera:2005tg} and the gaps, obtained from the  
minimization of (\ref{GKTP}) fulfill in general    
$\Delta_u \neq \Delta_d$.  
In the 2SC+X phase $\Delta_u=\Delta_d=\Delta$ and $a_{f,1}=a_{f,3}=1$,  
$a_{f,5}=(\Delta_X/\Delta)^2$ for $f=u, d$.
This simplifies Eq.~(\ref{integral1}) as long as no dependence of a density
 dependent strong coupling $\alpha_s$ is taken into account.
In the CSL-phase the functions for the several modes and flavors are
 introduced in Eq.~(\ref{En2}) with $p_d=xT+\mu_d,~~ p_u=yT+\mu_u$ and the
 dispersion relation $E_f=\sqrt{p_f^2+M_f^2}$.
The formulas are presented in a way to allow comparison with the
 spin-1 phase from the work of \cite{Schmitt:2005wg} (see Fig.~\ref{figs1}, lower panel).
The density dependent X-gap $\Delta_X$ was introduced in Ref. 
\cite{Grigorian:2004jq} for the first time to appropriately fit the cooling 
data of CS. 
Here we use the parametrization denoted as model IV in Ref. 
\cite{Popov:2005xa}, where $\Delta_X$ has been investigated more detailed to 
fulfill constraints from recent cooling phenomenology.   
The influence of the temperature dependence is taken into account by    
\begin{equation}
\label{tdepgap}  
\Delta(T)=\Delta_0\sqrt{1-(T/T_c)^{\beta}},  
\end{equation}  
where one can find values for $\beta$ between $1.0 - 3.2$ in the literature. 
In the following calculations we use $\beta=1.0$.

\begin{figure}[htp!]
\includegraphics[width=\columnwidth]{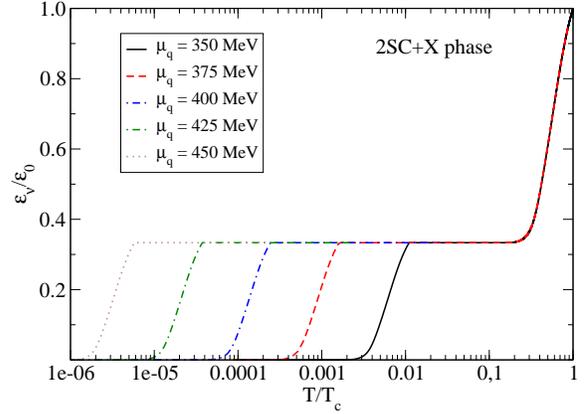}
\includegraphics[width=\columnwidth]{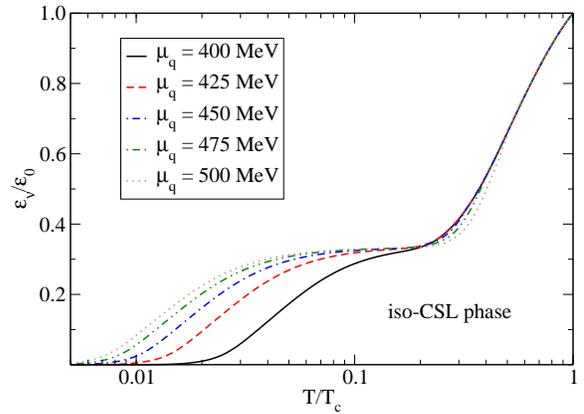}
\caption{Neutrino emissivities due to direct Urca processes in the 2SC+X 
phase (upper panel) and in the iso-CSL phase (lower panel). \label{figs1}}
\end{figure} 
In Figure \ref{figs1} we show the emissivities for the microscopic iso-CSL  
phase (lower panel) in comparison with the purely phenomenological 2SC+X phase 
(upper panel) as a function of temperature for different chemical potentials.  
For both phases a similar suppression of the emissivity is obtained. 
Hence the iso-CSL phase is probably able to explain recent cooling data in a  
more consistent way supporting the idea of superconducting phases in quark  
stars as explanation for observed fast CS cooling.    
  

\subsection{Bulk viscosity}  
  
According to \cite{Andersson:1997xt}, in the absence of viscosity all rotating
 CS would become unstable against r-modes  \cite{Andersson:2000mf}.  
Therefore, from the observation of millisecond pulsars, one can derive  
constraints for the composition of CS interiors  
\cite{Madsen:1999ci,Drago:2007iy}.   
For such an investigation, the bulk viscosity is a key quantity and we want to 
consider it for the two-flavor color superconducting phases introduced above, 
following the approach described in Ref. \cite{Sa'd:2006qv}. 
Note that the 2SC phase considered in \cite{Alford:2006gy}  is a three-flavor phase,  
where the nonleptonic process $u+d \leftrightarrow u+s$ provides the dominant  
contribution, see also \cite{Wang:2009if}. 
Due to absence of strange quarks in the 2SC phase of the  
present paper, this process does not occur.  \\
The bulk viscosity at all temperatures is determined by  
\begin{equation}  
\zeta=\frac{\lambda C_t^2}{\omega^2+(\lambda B/n)^2} ~, 
\end{equation}  
with $C_t=C+C^\prime$ and the coefficients functions  
\begin{eqnarray}  
C&=&\frac{M_u^2}{3\mu_u}-\frac{M_d^2}{3\mu_d},\nonumber\\  
C^\prime&=&\frac{4\alpha_s}{3\pi}\left[\frac{M_d^2}{\mu_d}\left({\rm ln}
\frac{2\mu_d}{M_d}-\frac{2}{3}\right)-\frac{M_u^2}{\mu_u}\left({\rm ln}
\frac{2\mu_u}{M_u}-\frac{2}{3}\right)\right],
\nonumber\\  
B&\simeq&\frac{\pi^2}{3}n\left(\frac{1}{\mu_u^2}+\frac{1}{\mu_d^2}+
\frac{1}{\mu_e^2}\right).  
\end{eqnarray}   

\begin{figure}[!htb]  
\includegraphics[width=\columnwidth]{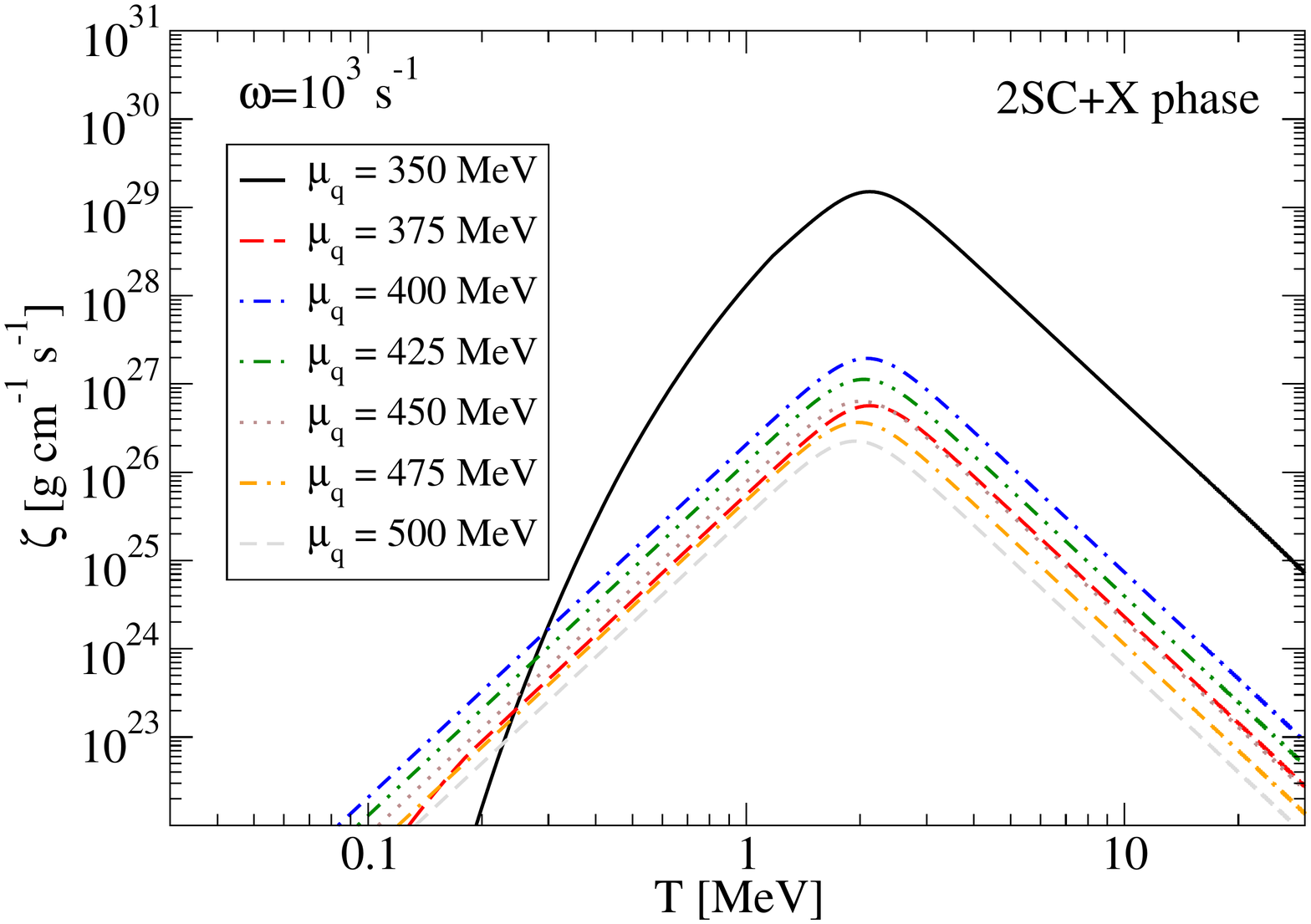}
\includegraphics[width=\columnwidth]{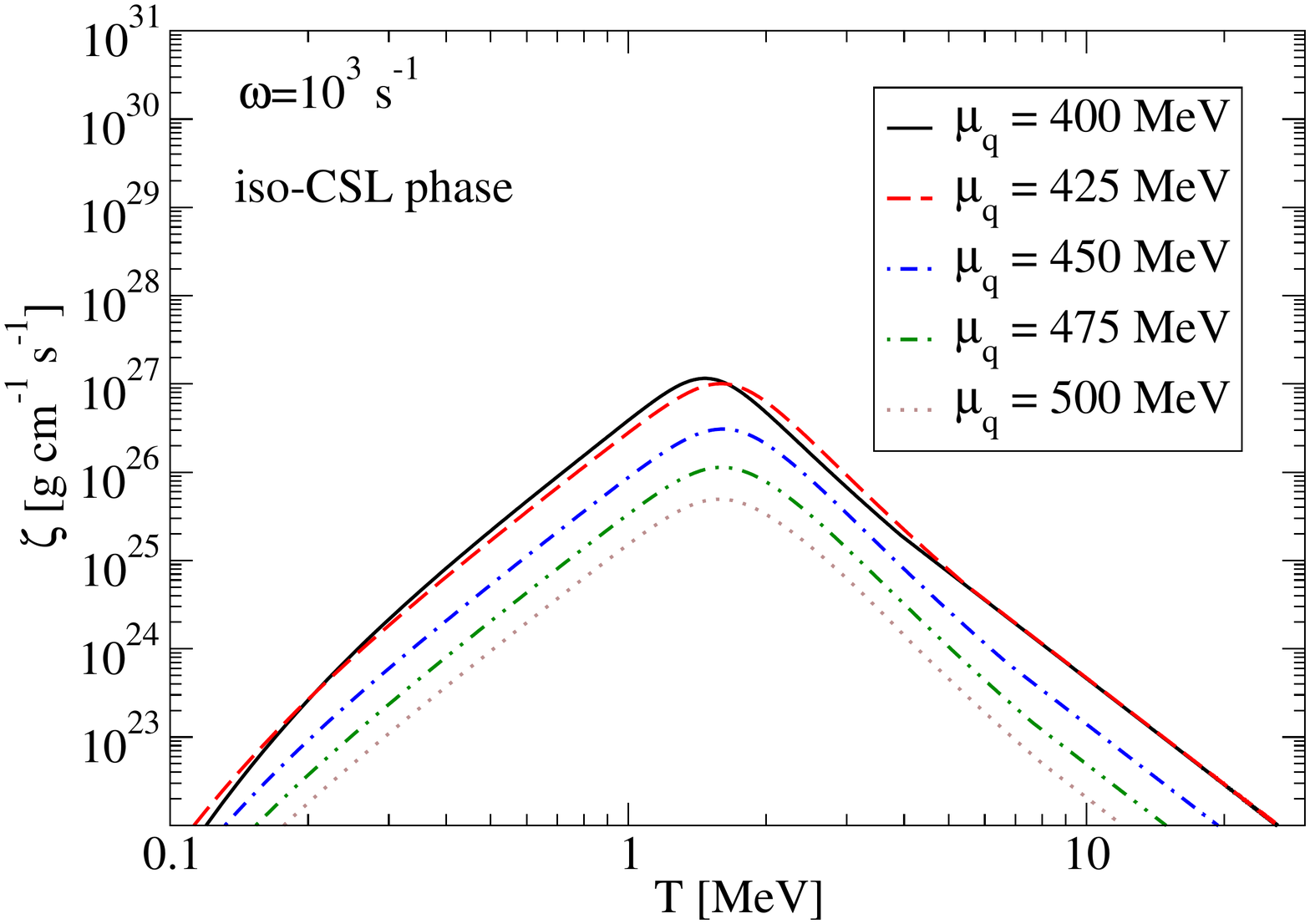}
\caption{Temperature dependence of bulk viscosity in the 2SC+X phase   
(upper panel) and in the iso-CSL phase (lower panel) for a frequency of $\omega=1$ kHz,  
typical for excitations of r-modes in millisecond pulsars.\label{viscosity}}
\end{figure}

The relevant processes for the bulk viscosity in two-flavor quark matter are  
the flavor changing weak processes of electron capture and beta decay, with a 
direct relation to the direct Urca emissivity 
\begin{eqnarray}  
\lambda&=&\frac{3}{2}\frac{\epsilon_0}{T^2}~ G_1(\Delta_u,\Delta_d)~.
\end{eqnarray}  
The numerical results for the NJL model in self-consistent meanfield  
approximation are displayed in Fig.~\ref{viscosity} for the 2SC+X phase   
(upper panel) and the iso-CSL phase (lower panel) in striking similarity.

Note that in comparison with Ref.~\cite{Sa'd:2006qv} the  peak  
value of the viscosity is also located at $T= 1 \sim 2$ MeV, but up to three 
orders of magnitude higher! Since the normal quark matter results coincide, 
this must be a result of the self-consistent treatment of masses, gaps and  
composition (chemical potentials) in the present models. 
In particular the strongly density dependent X-gap is rapidly decreasing with 
increasing density as one can see by the dramatic change for the bulk 
viscosity at low quark chemical potentials.
%

\section{Conclusions}
Transport properties in dense quark matter depend sensitively on the color
 superconductivity pairing patterns and provide thus a tool for unmasking
 the CS interiors by their cooling and rotational evolution characteristics. 
On the example of neutrino emissivities and bulk viscosities for the 2SC+X and
the iso-CSL phase we have demonstrated that both two-flavor color  
superconducting phases fulfil constraints from the CS phenomenology.   
For the 2SC+X phase with yet heuristic assumptions for the X-gap the hybrid 
star configurations and their cooling evolution have been numerically 
evaluated in accordance with observational data. 
The temperature and density behavior of the neutrino emissivity in the  
microscopically well-founded iso-CSL phase appear rather similar so that we  
expect a good agreement with CS cooling data too. 
The bulk viscosities for both phases have been presented here for the first  
time and provide sufficient damping of r-mode instabilities to comply with 
the phenomenology of rapidly spinning CS. 
We conclude that the subtle interplay between suppression of the direct Urca  
cooling process on the one hand and sufficiently large bulk viscosity puts  
severe constraints on microscopic approaches to quark matter in compact stars.



\section{Acknowledgements}

The authors acknowledge support from the Polish National Science Center (NCN) under grant numbers UMO-2011/02/A/ST2/00306 (A.K., D.B.) and UMO-2014/15/B/ST2/03752 (T.F.).
The work of D.B.  was supported in part  by the MEPhI Academic Excellence Project under contract number 02.a03.21.0005.

\begin{widetext}
\appendix
\section{Quark propagator}\label{B}
The free quark propagator $S_0(p)=({\not{p}-m})^{-1}$ can be represented as
\begin{eqnarray}\label{2.1}
S_0(p)&=&\frac{1}{\not{p}-m}=\frac{\not{p}+m}{(\not{p}-m)(\not{p}+m)}=
\frac{\not{p}+m}{p^2-m^2} 
\nonumber \\
S_0^{\pm}(p_0,{\bf p})&=&\frac{\not{p}+m}{(p_0\pm \mu)^2-{\bf p}^2-m^2}=
\frac{\not{p}+m}{(p_0\pm\mu)^2-E_p^2} 
\nonumber \\
&=& \frac{\not{p}+m}{[p_0\pm\mu-E_p][p_0\pm\mu+E_p]},
\end{eqnarray}
 with the particle energy $E_p=\sqrt{{\bf p}^2+m^2}$.
The upper and lower sign corresponds to particle and antiparticle contribution.
Eq.~(\ref{2.1}) can be extended to
\begin{eqnarray}\label{2.2}
S_0^{\pm}(p_0,{\bf p})&=& \frac{\not{p}+m}{[p_0\pm\mu-E_p][p_0\pm\mu+E_p]} ,
\nonumber\\ 
&=& \frac{1}{2E_p}\left(\frac{\not{p}+m}{p_0-E_p\pm\mu}-
\frac{\not{p}+m}{p_0+E_p\pm\mu}\right) ,
\nonumber \\
&=& \frac{1}{2E_p}\left(\frac{\gamma_0(p_0\pm\mu)-{\bf \gamma}{\bf p}+m}
{p_0-(E_p\mp\mu)}-\frac{\gamma_0(p_0\pm\mu)-{\bf \gamma}{\bf p}+m}
{p_0+(E_p\pm\mu)}\right), 
\nonumber \\
&=& \frac{1}{2E_p}\left(\frac{\gamma_0 E_p-{\bf \gamma}{\bf p}+m}
{p_0-E_p^{\mp}}+\frac{\gamma_0 E_p+{\bf \gamma}{\bf p}-m}{p_0+E_p^{\pm}} \right),
\end{eqnarray}
where the poles  $p^{\mp}_{0,a}=+(E_p\mp\mu)$ and 
$p_{0,b}^{\pm}=-(E_p\pm\mu)$ of the denominator have been replaced in the
 numerator. 
In addition we introduce the abbreviations $E_p^{\mp}=E_p\mp\mu$ and 
$E_p^{\pm}=E_p\pm\mu$ in the denominator for the corresponding particle/hole-
 and antiparticle/anti-hole excitation energies.
Simplifying one can write  
\begin{eqnarray} \label{2.3}
S_0^{\pm}(p_0,{\bf p}) &=& \frac{1}{2E_p}\left(\frac{\gamma_0 E_p-{\bf \gamma}
{\bf p}+m}{p_0-E_p^{\mp}}+\frac{\gamma_0 E_p+{\bf \gamma}{\bf p}-m}
{p_0+E_p^{\pm}}\right) 
\nonumber \\
&=& \frac{\frac{1}{2}\left(\gamma_0 -\frac{{\bf \gamma}{\bf p}}{E_p}+
\frac{m}{E_p}\right)}{p_0-E_p^{\mp}}+\frac{\frac{1}{2}\left(\gamma_0 +
\frac{{\bf \gamma}{\bf p}}{E_p}-\frac{m}{E_p}\right)}{p_0+E_p^{\pm}} 
\nonumber\\
&=& \frac{\gamma_0\frac{1}{2}\left(1 -\gamma_0{\bf \gamma}\hat{p}+\gamma_0
\hat{m}\right)}{p_0-E_p^{\mp}}+\frac{\gamma_0\frac{1}{2}\left(1 +\gamma_0{\bf 
\gamma}\hat{p}-\gamma_0 \hat{m}\right)}{p_0+E_p^{\pm}} 
\nonumber\\
&=& \frac{\gamma_0\frac{1}{2}[1 - \gamma_0 ({\bf \gamma}\hat{p}-\hat{m})]}
{p_0-E_p^{\mp}}-\frac{\gamma_0\frac{1}{2}[1 +\gamma_0({\bf \gamma}\hat{p}
-\hat{m})]}{p_0+E_p^{\pm}} 
\nonumber\\
&=& \frac{\gamma_0 \tilde{\Lambda}_p^-}{p_0-E_p^{\mp}}+\frac{\gamma_0 
\tilde{\Lambda}_p^+}{p_0+E_p^{\pm}},
\end{eqnarray}
where the energy projectors are of the form $\Lambda_p^{\pm}=\frac{1}{2}(1\pm\gamma_0~
 \mathcal{S}_p^{+})$,~$\tilde{\Lambda}_p^{\pm}=\frac{1}{2}(1\pm\gamma_0~ 
\mathcal{S}_p^{-})$ and $\mathcal{S}_p^{\pm}=\vec{\gamma}~\hat{p}\pm\hat{m}$ are
 introduced with their corresponding ``Foldy-Wouthhuysen'' matrices and 
$\hat{m}={m}/{E_p}$.  
The inverse free quark propagator is 
\begin{equation} \label{2.4}
[S_0^{\pm}]^{-1}=\gamma_0(p_0-E_p^{\mp})\Lambda_p^+ + \gamma_0(p_0+E_p^{\pm})
\Lambda_p^-.
\end{equation} 
To obtain the Nambu-Gorkov propagator we start with the identity
\begin{eqnarray} \label{3.1}
S^{-1}S&=&{\bf 1},\nonumber \\
\nonumber \\
\left( \begin{array}{cc}[S_0^+]^{-1} & \Delta^- \\ \Delta^+ & [S_0^-]^{-1} 
\end{array} \right) \left( \begin{array}{cc} A & B \\ C & D \end{array} 
\right) &=& \left( \begin{array}{cc} 1 & 0 \\ 0 & 1 \end{array} \right),
\end{eqnarray}
where we can find recursively the equations
\begin{eqnarray} 
\label{3.2}
{[S_0^+]}^{-1}~A+\Delta^-~C &=& 1, \nonumber\\
{[S_0^+]}^{-1}~B+\Delta^-~D &=& 0, \nonumber\\
\Delta^+~A+{[S_0^-]}^{-1}~C &=& 0, \nonumber\\
\Delta^+~B+{[S_0^-]}^{-1}~D &=& 1, 
\end{eqnarray} 
from which we derive the implicit expressions 
\begin{eqnarray} \label{3.3}
A &=& [(S_0^+)^{-1}-\Sigma^+]^{-1} = G^+ \nonumber\\
B &=& -S_0^+ \Delta^- G^- = F^- \nonumber\\
C &=& -S_0^- \Delta^+ G^+ = F^+ \nonumber\\
D &=& [(S_0^-)^{-1}-\Sigma^-]^{-1} = G^-, 
\end{eqnarray}
for the normal and anomalous parts of the Nambu-Gorkov propagator with the selfenergies
$\Sigma^{\pm}=\Delta^{\mp} S_0^{\mp} \Delta^{\pm}$ where the gap matrix for two-flavor quark matter 
are $\Delta^- = -i\Delta\varepsilon^{ik}\epsilon^{\alpha
\beta b}\gamma_5$ and $\Delta^+ = -i\Delta^{\ast}\varepsilon^{ik}
\epsilon^{\alpha\beta b}\gamma_5$.
The Nambu-Gorkov-Propagator obtains the form
\begin{equation} \label{3.4}
S = \left(\begin{array}{cc} G^+ & F^- \\ F^+ & G^- \end{array} \right),
\end{equation}
with the normal parts 
\begin{eqnarray} \label{3.5}
G^{\pm}&=&[(S_0^{\pm})^{-1}-\Delta^{\mp} S_0^{\mp} \Delta^{\pm}]^{-1}
=\frac{p_0+E_p^{\mp}}{p_0^2-(\xi_p^{\mp})^2}\gamma_0\tilde{\Lambda}_p^-+
\frac{p_0-E_p^{\pm}}{p_0^2-(\xi_p^{\pm})^2}\gamma_0\tilde{\Lambda}_p^+,
\end{eqnarray}
and the anomalous parts 
\begin{equation} \label{3.6}
F^{\pm}=-S_0^{\mp}\Delta^{\pm}G^{\pm}
= \frac{\Delta^{\pm}}{p_0^2-(\xi_p^{\pm})^2}\tilde{\Lambda}_p^++\frac{
\Delta^{\pm}}{p_0^2-(\xi_p^{\mp})^2}\tilde{\Lambda}_p^-.
\end{equation}
The four poles of the Nambu-Gorkov propagators $p_0=\pm \xi_p^-$ and 
$p_0=\mp \xi_p^+$ with $(\xi_p^{\pm})^2=(E_p^{\pm})^2+\Delta^2$ correspond to
 the quasi-particle/quasi-hole and quasi-antiparticle/quasi-anti-hole
 excitation energy in the color superconducting phase.

\section{Polarization tensor}\label{C}
\begin{center}
\begin{figure}[h!]
\includegraphics[width=0.9\columnwidth]{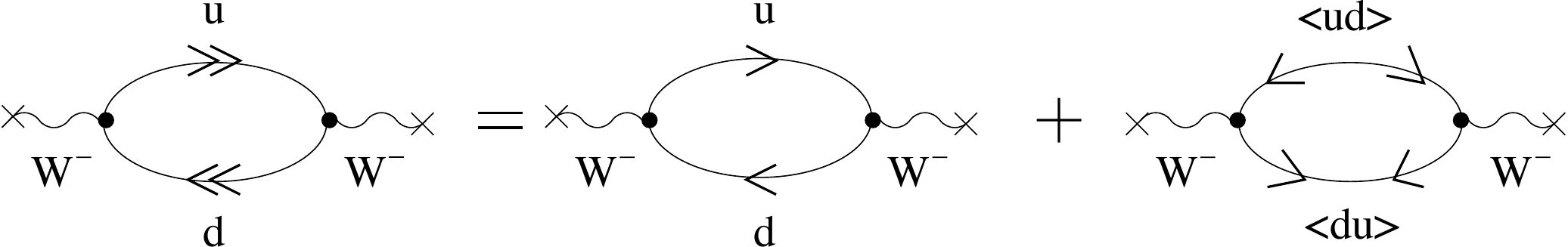}
\caption{{\small Decomposition of the polarization loop diagram for the hadronic tensor into normal and anomalous parts.}
\label{b1.1}}
\end{figure}
The decomposition of the polarization loop diagram for the hadronic tensor in superconducting quark matter as shown in Fig.~\ref{b1.1} has the following Nambu Gorkov structure
\end{center}
\begin{equation} \label{4.1}
\Pi_{\mu\nu}(q)=-\frac{i}{2}\int\frac{d^4 p}{(2\pi)^4}{\rm Tr_{Z,D}}~[\Gamma_{
\mu}^Z~ S_p ~\Gamma_{\nu}^Z~ S_{p+q}],
\end{equation}
with $\Gamma_i^Z=\left(\begin{array}{cc} \Gamma_i^- & 0 \\ 0 & \Gamma_i^+ 
\end{array}\right)$, where $\Gamma_i^{\pm}=\gamma_{\mu}(\mathcal{V} \pm 
\mathcal{A}\gamma_5)$ for $i=\mu,\nu$ and $S_j=\left( \begin{array}{cc} G_j^+
 & F_j^- \\ F_j^+ & G_j^- \end{array} \right)$ for $j=p,p+q$.\\
The vector and axial vector couplings of charged quark currents are
 $\mathcal{V}=\cos \theta_c$ and  $\mathcal{A}=\cos \theta_c$ and the vertices
 have therefore the form $\Gamma_i^{\pm}= \cos \theta_c \gamma_{\mu}(1 \pm 
\gamma_5)$.
Performing the trace over the Nambu-Gorkov space leads to
\begin{equation}  \label{4.2}
\Pi_{\mu\nu}(q)= -\frac{i}{2}\cos^2\theta_c \int \frac{d^4 p}{(2\pi)^4} 
{\rm Tr_D}~[\Gamma_{\mu}^- G_p^+ \Gamma_{\nu}^- G_{p+q}^+ + \Gamma_{\mu}^+ 
G_p^- \Gamma_{\nu}^+ G_{p+q}^- + \Gamma_{\mu}^- F_p^- \Gamma_{\nu}^+ F_{p+q}^+
 + \Gamma_{\mu}^+ F_p^+ \Gamma_{\nu}^- F_{p+q}^- ]~,
\end{equation}
which can be expressed by 
\begin{equation} 
\label{4.3}
\Pi_{\mu\nu}(q_0,{\bf q})= -i\frac{T}{2} \cos^2\theta_c\sum_n \int \frac{d^3 
{\bf p}}{(2\pi)^3} {\rm Tr_D}~[\Gamma_{\mu}^- G_p^+ \Gamma_{\nu}^- G_{p+q}^+ +
 \Gamma_{\mu}^+ G_p^- \Gamma_{\nu}^+ G_{p+q}^- + \Gamma_{\mu}^- F_p^- 
\Gamma_{\nu}^+ F_{p+q}^+ + \Gamma_{\mu}^+ F_p^+ \Gamma_{\nu}^- F_{p+q}^- ] 
\end{equation}
as summation over fermionic ($p_0=i(2n+1)\pi T$) and bosonic ($q_0=i2m\pi T$) Matsubara frequencies.
The trace over the Dirac space indices results in
\begin{eqnarray} 
\label{4.4}
&&{\rm Tr_D}[\Gamma_{\mu}^- G_p^+ \Gamma_{\nu}^- G_{p+q}^+ + \Gamma_{\mu}^+ 
G_p^- \Gamma_{\nu}^+ G_{p+q}^- + \Gamma_{\mu}^- F_p^- \Gamma_{\nu}^+ F_{p+q}^+
 + \Gamma_{\mu}^+ F_p^+ \Gamma_{\nu}^- F_{p+q}^- ] = 
\nonumber\\
&=&\sum\limits_r\Biggl( \frac{(p_0+E_p^-)(p_0+q_0+E_k^-)}{[p_0^2-\xi_{p,r}^2]
[(p_0+q_0)^2-\xi_{k,r}^2]} 
\Bigl\{\mathcal{T}_{\mu\nu}^+(\hat{p},\hat{k})+\widetilde{\mathcal{T}}_{\mu
\nu}^+(\hat{p},\hat{k})-\left[\widetilde{\mathcal{W}}_{\mu\nu}^+(\hat{p},
\hat{k})+\mathcal{W}_{\mu\nu}^+(\hat{p},\hat{k})\right]\Bigr\}  
\nonumber\\
&+&\frac{(p_0-E_p^-)(p_0+q_0-E_{k}^-)}{[p_0^2-\xi_{p,r}^2][(p_0+q_0)^2
-\xi_{k,r}^2]}
\Bigl\{\mathcal{T}_{\mu\nu}^-(\hat{p},\hat{k})+\widetilde{\mathcal{T}}_{\mu
\nu}^-(\hat{p},\hat{k})+\left[\widetilde{\mathcal{W}}_{\mu\nu}^-(\hat{p},
\hat{k})+\mathcal{W}_{\mu\nu}^-(\hat{p},\hat{k})\right]\Bigr\} 
\nonumber\\
&-& \frac{\Delta^2}{[p_0^2-\xi_{p,r}^2][(p_0+q_0)^2-\xi_{k,r}^2]}
\Bigl\{\mathcal{T}_{\mu\nu}^-(\hat{p},\hat{k})+\mathcal{T}_{\mu\nu}^+(\hat{p},
\hat{k})+\widetilde{\mathcal{T}}_{\mu\nu}^-(\hat{p},\hat{k})+\widetilde{
\mathcal{T}}_{\mu\nu}^+(\hat{p},\hat{k})
\nonumber\\
 &-&  \left[ \widetilde{\mathcal{W}}_{\mu\nu}^+(\hat{p},\hat{k})+\mathcal{W}_{\mu
\nu}^+(\hat{p},\hat{k})
-\widetilde{\mathcal{W}}_{\mu\nu}^-(\hat{p},\hat{k})-\mathcal{W}_{\mu\nu}^-
 (\hat{p},\hat{k})\right] \Bigr\}\Biggr) 
\nonumber \\
\end{eqnarray}
with the tensors $\mathcal{T}_{\mu\nu}^{\pm}(\hat{p},\hat{k})=
{\rm Tr}[\gamma_0\gamma_{\mu}\Lambda_p^{\pm}\gamma_0\gamma_{\nu}
\Lambda_k^{\pm}]$, $\mathcal{W}_{\mu\nu}^{\pm}(\hat{p},\hat{k})=
{\rm Tr}[\gamma_0\gamma_{\mu}\Lambda^{\pm}_p\gamma_0\gamma_{\nu}
\Lambda_k^{\pm}\gamma_5]$, $\widetilde{\mathcal{T}}_{\mu\nu}^{\pm}
(\hat{p},\hat{k})={\rm Tr}[\gamma_0\gamma_{\mu}\widetilde{\Lambda}_p^{\pm}
\gamma_0\gamma_{\nu}\Lambda_k^{\pm}]$ and $\widetilde{\mathcal{W}}_{\mu\nu}^{
\pm}(\hat{p},\hat{k})={\rm Tr}[\gamma_0\gamma_{\mu}\widetilde{\Lambda}^{\pm}_p
\gamma_0\gamma_{\nu}\Lambda_k^{\pm}\gamma_5]$.
\\ 
We introduce the notation $k=p+q$ for simplification.
Performing the Matsubara summation results in the expressions
\begin{eqnarray} \label{4.6}
  A^{\pm}( E_p, E_k) &=& T\sum_n\frac{(p_0\pm E_p^-)(p_0+q_0\pm E_k^-)}{
[p_0^2-(\xi_p^{-})^2][(p_0+q_0)^2-(\xi_{k}^{-})^2]}=-\frac{1}{2 \xi_p^- 2
\xi_k^-}\sum_{s_1s_2=\pm}\frac{(\xi_p^-+s_1E_p^-)(\xi_k^-+s_2E_k^-)}{q_0\pm
 s_1\xi_p^-\mp s_2\xi_k^-}\frac{n_F(\pm s_1\xi_p^-)n_F(\mp s_2\xi_k^-)}{n_B(
\pm s_1\xi_p^-\mp s_2\xi_k^-)} 
\nonumber\\
B(E_p,E_k) &=& T\sum_n\frac{\Delta^2}{[p_0^2-(\xi_p^{-})^2][(p_0+q_0)^2
-(\xi_{k}^{-})^2]}=-\frac{1}{2 \xi_p^- 2\xi_k^-}\sum_{s_1s_2=\pm}
\frac{1}{q_0+ s_1\xi_p^-- s_2\xi_k^-}\frac{n_F(s_1\xi_p^-)n_F(-s_2\xi_k^-)}{
n_B(s_1\xi_p^-- s_2\xi_k^-)}
\nonumber\\
\end{eqnarray}
so that we obtain for the polarization loop   
\begin{eqnarray} \label{4.5}
\Pi_{\mu\nu}(q_0,{\bf q})&=&-i\sum\limits_r\int\frac{{\rm d}^3 {\bf p}}{(2
\pi)^3}A_r^+(E_p,E_k)\{\mathcal{T}_{\mu\nu}^+(\hat{p},\hat{k})+
\widetilde{\mathcal{T}}_{\mu\nu}^+(\hat{p},\hat{k})-[\widetilde{\mathcal{W}}_{
\mu\nu}^+(\hat{p},\hat{k})+\mathcal{W}_{\mu\nu}^+(\hat{p},\hat{k})]\}
\nonumber\\
&~&~~~~+A_r^-(E_p,E_k)\{\mathcal{T}_{\mu\nu}^-(\hat{p},\hat{k})+\widetilde{
\mathcal{T}}_{\mu\nu}^-(\hat{p},\hat{k})+[\widetilde{\mathcal{W}}_{\mu\nu}^-
(\hat{p},\hat{k})+\mathcal{W}_{\mu\nu}^-(\hat{p},\hat{k})]\}
\nonumber\\
&~&~~~~-\Delta^2B_r(E_p,E_k)\{\mathcal{T}_{\mu\nu}^-(\hat{p},\hat{k})+
\mathcal{T}_{\mu\nu}^+(\hat{p},\hat{k})+\widetilde{\mathcal{T}}_{\mu\nu}^-(
\hat{p},\hat{k})+\widetilde{\mathcal{T}}_{\mu\nu}^+(\hat{p},\hat{k})
\nonumber\\
&~&~~~~-[\widetilde{\mathcal{W}}_{\mu\nu}^+(\hat{p},\hat{k})+\mathcal{W}_{\mu
\nu}^+(\hat{p},\hat{k})-\widetilde{\mathcal{W}}_{\mu\nu}^-(\hat{p},\hat{k})
-\mathcal{W}_{\mu\nu}^-(\hat{p},\hat{k})]\},
\end{eqnarray}
with the corresponding abbreviations
\begin{eqnarray} 
\label{4.6}
  A_r^{\pm}(E_p,E_k) &=\left[\frac{B_p^{\mp}B_k^{\mp}}{q_0-\xi_{p,r}
+\xi_{k,r}}-\frac{B_p^{\pm}B_k^{\pm}}{q_0+\xi_{p,r}-\xi_{k,r}}\right]
\frac{n_F(\xi_{p,r})n_F(-\xi_{k,r})}{n_B(\xi_{p,r}-\xi_{k,r})} 
\nonumber \\
&~~~~~~~+\left[\frac{B_p^{\mp}B_k^{\pm}}{q_0-\xi_{p,r}-\xi_{k,r}}
-\frac{B_p^{\pm}B_k^{\mp}}{q_0+\xi_{p,r}+\xi_{k,r}}\right]\frac{n_F(\xi_{p,r})
n_F(\xi_{k,r})}{n_B(\xi_{p,r}+\xi_{k,r})}
\nonumber\\
 B_r(E_p,E_k)&=\frac{1}{4\xi_{p,r}\xi_{k,r}}\left[\frac{1}{q_0-\xi_{p,r}
+\xi_{k,r}}-\frac{1}{q_0+\xi_{p,r}-\xi_{k,r}}\right]\frac{n_F(\xi_{p,r})
n_F(-\xi_{k,r})}{n_B(\xi_{p,r}-\xi_{k,r})} 
\nonumber \\
&~~~~~~~+\left[\frac{1}{q_0+\xi_{p,r}+\xi_{k,r}}-\frac{1}{q_0-\xi_{p,r}
-\xi_{k,r}}\right]\frac{n_F(\xi_{p,r})n_F(\xi_{k,r})}{n_B(\xi_{p,r}
+\xi_{k,r})}.
\end{eqnarray}
and the Bogoliubov coefficients
\begin{equation}
B^e_i\equiv \frac{\xi_{i,r}+eE_i^-}{2\xi_{i,r}},~~~~(i=p,k;~~e=\pm).
\end{equation}
The first term and the second term on the right hand side of Eq.~(\ref{4.5}) provide the same
 contribution, since the second term is the charge conjugated counterpart of the first one. 
\begin{eqnarray} 
\label{4.9a}
\Pi_{\mu\nu}(q_0,{\bf q})&=&-i\sum\limits_r\int\frac{{\rm d}^3 {\bf p}}{
(2\pi)^3}\biggl(2 A_r(E_p,E_k)\{\mathcal{T}_{\mu\nu}^+(\hat{p},\hat{k})+
\widetilde{\mathcal{T}}_{\mu\nu}^+(\hat{p},\hat{k})-[\widetilde{\mathcal{W}}_{
\mu\nu}^+(\hat{p},\hat{k})+\mathcal{W}_{\mu\nu}^+(\hat{p},\hat{k})]\}
\nonumber\\
&~&-\Delta^2~B_r(E_p,E_k)\{\mathcal{T}_{\mu\nu}^-(\hat{p},\hat{k})+
\mathcal{T}_{\mu\nu}^+(\hat{p},\hat{k})+\widetilde{\mathcal{T}}_{\mu\nu}^-(
\hat{p},\hat{k})+\widetilde{\mathcal{T}}_{\mu\nu}^+(\hat{p},\hat{k})
\nonumber\\
&~&-[\widetilde{\mathcal{W}}_{\mu\nu}^+(\hat{p},\hat{k})
+\mathcal{W}_{\mu\nu}^+(\hat{p},\hat{k})-\widetilde{\mathcal{W}}_{\mu\nu}^-(
\hat{p},\hat{k})-\mathcal{W}_{\mu\nu}^-(\hat{p},\hat{k})]\}\biggl).
\end{eqnarray}
The hadronic tensors $\mathcal{W}^{\pm}_{\mu\nu}(\hat{p},\hat{k})$
 and $\mathcal{\widetilde{W}}^{\pm}_{\mu\nu}(\hat{p},\hat{k})$ are identical
 (Appendix \ref{D}) and the expression for the polarization tensor becomes
\begin{eqnarray} \label{4.9b} 
\Pi_{\mu\nu}(q_0,{\bf q})&=&-i\sum\limits_r\int\frac{{\rm d}^3 {\bf p}}{
(2\pi)^3}\bigg(2 A_r(E_p,E_k)[\mathcal{T}_{\mu\nu}^+(\hat{p},\hat{k})+
\widetilde{\mathcal{T}}_{\mu\nu}^+(\hat{p},\hat{k})-2\mathcal{W}_{\mu\nu}^+
(\hat{p},\hat{k})]
\nonumber\\
&~& -\Delta^2~B_r(E_p,E_k)\{\mathcal{T}_{\mu\nu}^-(\hat{p},\hat{k})+
\mathcal{T}_{\mu\nu}^+(\hat{p},\hat{k})+\widetilde{\mathcal{T}}_{\mu\nu}^-
(\hat{p},\hat{k})+\widetilde{\mathcal{T}}_{\mu\nu}^+(\hat{p},\hat{k})
-2[\mathcal{W}_{\mu\nu}^+(\hat{p},\hat{k})-\mathcal{W}_{\mu\nu}^-(\hat{p},
\hat{k})]\}\bigg).
\end{eqnarray}
By analytic expansion and use of the Dirac identity $$\frac{1}{
(x+i\eta)^{n+1}}=\mathcal{P}\frac{1}{x^{n+1}}-i\pi\frac{(-1)^n}{n!}
\delta^{(n)}(x)$$ one can extract the imaginary part of the polarization tensor
\begin{equation} \label{4.10}
{\rm Im} \Pi^R_{\mu\nu}(q_0,{\bf q})=-\pi\cos^2\theta_c\sum\limits_r\int
\frac{{\rm d}^3 {\bf p}}{(2\pi)^3}\biggl(2~A_r^{\ast}(E_p,E_k)\mathcal{H}_{\mu
\nu}^{(\rm n)}(\hat{p},\hat{k})-\Delta^2~B_r^{\ast}(E_p,E_k)\mathcal{H}_{\mu
\nu}^{(\rm a)}(\hat{p},\hat{k})\biggl),
\end{equation}
where
\begin{eqnarray}
  A_r^{\ast}(E_p,E_k) &= \sum\limits_{e_1,e_2=\pm}B_p^{e_1}B_k^{e_2}
\frac{n_F(-e_1\xi_{p,r})n_F(e_2\xi_{k,r})}{n_B(-e_1\xi_{p,r}+e_2\xi_{k,r})}
\delta(q_0+e_1\xi_{p,r}-e_2\xi_{k,r})
\nonumber\\
B_r^{\ast}(E_p,E_k) &= \sum\limits_{e_1,e_2=\pm}\frac{e_1e_2}{4\xi_{p,r}
\xi_{k,r}}
\frac{n_F(e_1\xi_{p,r})n_F(-e_2\xi_{k,r})}{n_B(e_1\xi_{p,r}-e_2\xi_{k,r})}
\delta(q_0-e_1\xi_{p,r}+e_2\xi_{k,r})~.
\end{eqnarray}
The normal and anomalous hadronic tensors are
\begin{eqnarray} \label{4.12}
\mathcal{H}_{\mu\nu}^{(\rm n)}&=&\mathcal{T}_{\mu\nu}^+(\hat{p},\hat{k})+
\widetilde{\mathcal{T}}_{\mu\nu}^+(\hat{p},\hat{k})-2\mathcal{W}_{\mu\nu}^+
(\hat{p},\hat{k})\\
\mathcal{H}_{\mu\nu}^{(\rm a)}&=&\mathcal{T}_{\mu\nu}^-(\hat{p},\hat{k})+
\mathcal{T}_{\mu\nu}^+(\hat{p},\hat{k})+\widetilde{\mathcal{T}}_{\mu\nu}^-
(\hat{p},\hat{k})+\widetilde{\mathcal{T}}_{\mu\nu}^+(\hat{p},\hat{k})
-2[\mathcal{W}_{\mu\nu}^+(\hat{p},\hat{k})-\mathcal{W}_{\mu\nu}^-(\hat{p},
\hat{k})].
\end{eqnarray}

\section{Kinematics}\label{D}

\begin{figure}[h!]
\includegraphics[width=0.5\textwidth]{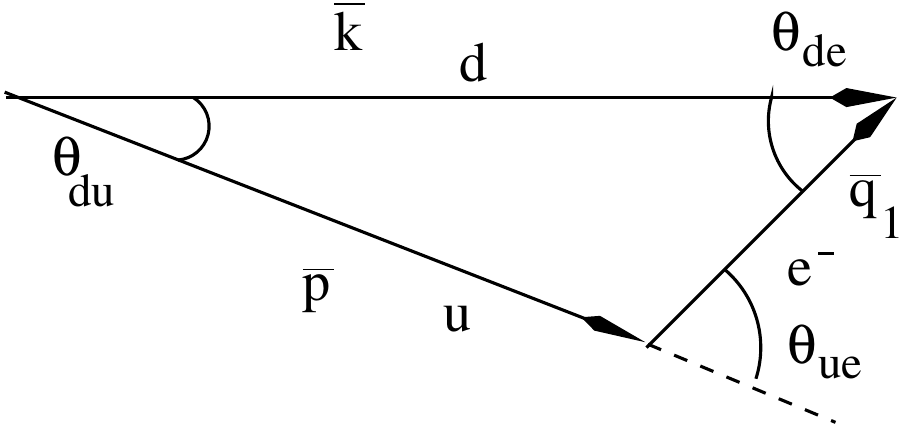}
\caption{Triangle of momentum conservation for the direct Urca process}
\label{b1.3}
\end{figure}

From Fig. \ref{b1.3} one finds the relations
\begin{eqnarray} 
\label{5.1}
p_{F,u}\cos \theta_{du}+p_{F,e} \cos \theta_{de} &=& p_{F,d}, \\
p_{F,u}\sin\theta_{du} &=& p_{F,e}\sin\theta_{de}.
\end{eqnarray}
The cosine and sinus of the angles $\theta_{du}$ and $\theta_{de} \ll 1$ can be
 expanded in a power series to second order
\begin{eqnarray} 
\label{5.2a}
p_{F,d}-p_{F,u}-p_{F,e}&=&-\frac{1}{2}(p_{F,u}~\theta^2_{du}+p_{F,e}~\theta_{de}^2),\\
p_{F,u}~\theta_{du}&=&p_{F,e}~\theta_{de}~.
\label{5.2b}
\end{eqnarray}
Eliminating $\theta_{du}$ by inserting (\ref{5.2b}) to (\ref{5.2a}) one gets
in lowest order of $\theta_{de}$ and $p_{F,e}/p_{F,u}$ 
\begin{eqnarray} 
\label{5.3}
p_{F,d}-p_{F,u}-p_{F,e} \simeq -\frac{1}{2}p_{F,e}~
\theta_{de}^2\left(1+\frac{p_{F,e}}{p_{F,u}}\right)
\simeq -\frac{1}{2}p_{F,e}~\theta_{de}^2, 
\end{eqnarray}
which corresponds to the momentum conservation of the direct Urca process.
Note that $\theta_{de}\simeq \theta_{ue}$. 
Momentum conservation ensures that all particles are collinear, as long as quark masses are neglected 
or no perturbative corrections in the dispersion relation of the free quarks are included.
Interactions modify the dispersion relation and lead to a nonvanishing matrix element. 
Quark-quark interactions can be treated either perturbatively in lowest order of the strong coupling constant $\alpha_s$
\begin{equation} 
\label{5.4}
\mu_i\simeq p_{F,i}\left(1+\frac{2}{3\pi}\alpha_s\right),~~~~~~\alpha_s=
\frac{g^2}{4\pi},~~~~i=u,d
\end{equation}
or due to the effect of finite quark masses
\begin{equation} 
\label{5.6}
\mu_i=\sqrt{p_{F,i}^2+m_i^2} \simeq  p_{F,i}\left[1+\frac{1}{2}
\left(\frac{m_i}{p_{F,i}}\right)^2\right],~~~~~~~~~i=u,d,e .
\end{equation}
Using Eqs.~(\ref{5.4}), (\ref{5.6}) and the $\beta$-equilibrium condition
$\mu_d=\mu_u+\mu_e$ one can find the equivalent expressions for the angle
 $\theta_{de}$ corresponding to Eq.~(\ref{5.3})
\begin{eqnarray}
\label{5.7}
\theta_{de}^2 \simeq \left\{ \begin{array}{cl} 
  \frac{4}{3\pi}\alpha_s \\[0.2cm]
\frac{m_d^2}{p_{F,e}p_{F,d}}
\left[1-\left(\frac{m_u}{m_d}\right)^2\left(\frac{p_{F,d}}{p_{F,u}}\right)
-\left(\frac{m_e}{m_d}\right)^2\left(\frac{p_{F,d}}{p_{F,e}}\right)\right]
\\
\end{array}\right.,
\end{eqnarray}
which is crucial to calculate the emissivity, see also \cite{Iwamoto:1982}.

\section{Contractions} {\label F}

The contraction of the matrix elements of the leptonic tensor (\ref{Ltensor}) gives 
\begin{eqnarray}
\label{D1}
\mathcal{L}^{00}(q_1,q_2)&=&8(q_1^0q_2^0+{\bf q_1 \cdot q_2})~,
\nonumber\\
\mathcal{L}^{0i}(q_1,q_2)&=&8[q_1^0q_2^i+ q_1^iq_2^0-i
\epsilon^{ijk}q_{1j}q_{2k}]~,
\nonumber\\
\mathcal{L}^{i0}(q_1,q_2)&=&8[q_1^0q_2^i+ q_1^iq_2^0+i\epsilon^{ijk}q_{1j} q_{2k}]~,
\nonumber\\
\mathcal{L}^{ij}(q_1,q_2)&=&8[\delta^{ij}(q_1^0q_2^0-{\bf q_1 \cdot q_2})
+ q_1^i q_2^j+ q_1^j q_2^i-i\epsilon^{ijkl}q_{1k}q_{2l}]~. 
\end{eqnarray}
The elements of the tensors $\mathcal{T}_{\mu\nu}^{\pm}(\hat{p},\hat{k})$,
$\widetilde{\mathcal{T}}_{\mu\nu}^{\pm}(\hat{p},\hat{k})$,
$\mathcal{W}_{\mu\nu}^{\pm}(\hat{p},\hat{k})$,
$\widetilde{\mathcal{W}}_{\mu\nu}^{\pm}(\hat{p},\hat{k})$
in Eq.~(\ref{4.4}) become
\begin{alignat}{4}
\mathcal{T}_{00}^{\pm}(\hat{p},\hat{k}) & = 1+\hat{p}\cdot \hat{k}+\hat{m}_u\hat{m}_d ~, 
& \qquad  \mathcal{\widetilde{T}}_{00}^{\pm}(\hat{p},\hat{k})
& = 1+\hat{p}\cdot \hat{k}-\hat{m}_u\hat{m}_d~, & 
\nonumber\\
\mathcal{T}_{0i}^{\pm}(\hat{p},\hat{k}) & = \pm(\hat{p}_i+\hat{k}_i)~, 
& \qquad \mathcal{\widetilde{T}}_{0i}^{\pm}(\hat{p},\hat{k}) & = \pm(\hat{p}_i +\hat{k}_i)~, 
\nonumber\\
\mathcal{T}_{i0}^{\pm}(\hat{p},\hat{k})& = \pm(\hat{p}_i+\hat{k}_i)~, 
& \qquad \mathcal{\widetilde{T}}_{i0}^{\pm}(\hat{p},\hat{k})
& = \pm(\hat{p}_i+\hat{k}_i)~, 
\nonumber\\
\mathcal{T}_{ij}^{\pm}(\hat{p},\hat{k})&=  \delta_{ij}(1-\hat{p}\cdot \hat{k}
-\hat{m}_u\hat{m}_d)+\hat{p}_i\hat{k}_j+\hat{k}_i\hat{p}_j~, 
& \qquad \mathcal{\widetilde{T}}_{ij}^{\pm}(\hat{p},\hat{k})& = \delta_{ij}(1-\hat{p}
\cdot \hat{k}+\hat{m}_u\hat{m}_d)+\hat{p}_i\hat{k}_j+\hat{k}_i\hat{p}_j~. 
\label{D2}
\end{alignat}

\begin{eqnarray}
\label{D3}
\mathcal{W}_{00}^{\pm}(\hat{p},\hat{k})&=& \mathcal{\widetilde{W}}_{00}^{\pm} (\hat{p},\hat{k}) = 0~,
\nonumber\\
\mathcal{W}_{0i}^{\pm}(\hat{p},\hat{k})&=& \mathcal{\widetilde{W}}_{0i}^{\pm}
(\hat{p},\hat{k})= -i\epsilon_{ijk}\hat{p}^j\hat{k}^k~,
\nonumber\\
\mathcal{W}_{i0}^{\pm}(\hat{p},\hat{k})&=&\mathcal{\widetilde{W}}_{i0}^{\pm}
(\hat{p},\hat{k})= +i\epsilon_{ijk}\hat{p}^j\hat{k}^k~,
\nonumber\\
\mathcal{W}_{ij}^{\pm}(\hat{p},\hat{k})&=& \mathcal{\widetilde{W}}_{ij}^{\pm}
(\hat{p},\hat{k})= \mp i\epsilon_{ijk}(\hat{p}^k-\hat{k}^k)+i\epsilon_{ijkl} \hat{p}^k\hat{k}^l~.
\end{eqnarray}
With (\ref{D1}), (\ref{D2}) and (\ref{D3}) we obtain for the Lorentz contraction
\begin{eqnarray}
\mathcal{L}^{\mu\nu}(q_1,q_2)\mathcal{T}_{\mu\nu}^{\pm}(\hat{p},\hat{k})
&=&16[(q_1^0\mp{\bf q_1}\cdot \hat{p})(q_2^0\mp{\bf q_2}\cdot \hat{k})
+(q_1^0\mp{\bf q_1}\cdot\hat{k})(q_2^0\mp{\bf q_2}\cdot \hat{p})-(q_1^0q_2^0
-{\bf q_1 \cdot q_2})\hat{m}_u\hat{m}_d]~,
\nonumber\\
\mathcal{L}^{\mu\nu}(q_1,q_2)\mathcal{\widetilde{T}}_{\mu\nu}^{\pm}(\hat{p},
\hat{k})&=&16[(q_1^0\mp{\bf q_1}\cdot \hat{p})(q_2^0\mp{\bf q_2}\cdot \hat{k})
+(q_1^0\mp{\bf q_1}\cdot\hat{k})(q_2^0\mp{\bf q_2}\cdot \hat{p})+(q_1^0q_2^0
-{\bf q_1 \cdot q_2})\hat{m}_u\hat{m}_d]~,
\nonumber\\
\mathcal{L}^{\mu\nu}(q_1,q_2)\mathcal{W}_{\mu\nu}^{\pm}(\hat{p},\hat{k})
&=& 16[(q_1^0\mp{\bf q_1}\cdot\hat{k})(q_2^0\mp{\bf q_2}\cdot\hat{p})
-(q_1^0\mp{\bf q_1}\cdot\hat{p})(q_2^0\mp{\bf q_2}\cdot\hat{k})]~.
\end{eqnarray}

Contraction between leptonic $\mathcal{L}^{\mu\nu}(q_1,q_2)$ and hadronic tensor of
 the  normal phase $\mathcal{H}_{\mu\nu}^{(\rm n)}(\hat{p},\hat{k})$ results in
\begin{eqnarray}
\mathcal{L}^{\mu\nu}(q_1,q_2)\mathcal{H}_{\mu\nu}^{(\rm n)}(\hat{p},\hat{k}) 
&=& \mathcal{L}^{\mu\nu}(q_1,q_2)\mathcal{T}_{\mu\nu}^{+}(\hat{p},\hat{k})+
\mathcal{L}^{\mu\nu}(q_1,q_2)\mathcal{\widetilde{T}}_{\mu\nu}^{+}(\hat{p},
\hat{k})-\mathcal{L}^{\mu\nu}(q_1,q_2)\mathcal{W}_{\mu\nu}^+
\nonumber\\
&=& 16[(q_1^0-{\bf q_1}\cdot\hat{p})(q_2^0-{\bf q_2}\cdot\hat{k})+(q_1^0-
{\bf q_1}\cdot\hat{k})(q_2^0-{\bf q_2}\cdot\hat{p})
\nonumber\\
&~&-(q_1^0q_2^0-{\bf q_1}\cdot{\bf q_2})\hat{m_u}\hat{m_d}
\nonumber\\
&~&+(q_1^0-{\bf q_1}\cdot\hat{p})(q_2^0-{\bf q_2}\cdot\hat{k})+(q_1^0
-{\bf q_1}\cdot\hat{k})(q_2^0-{\bf q_2}\cdot\hat{p})
\nonumber\\
&~&+(q_1^0q_2^0-{\bf q_1}\cdot{\bf q_2})\hat{m_u}\hat{m_d}
\nonumber\\
&~&-2(q_1^0-{\bf q_1}\cdot\hat{k})(q_2^0-{\bf q_2}\cdot\hat{p})+2(q_1^0
-{\bf q_1}\cdot\hat{p})(q_2^0-{\bf q_2}\cdot\hat{k})]
\nonumber\\
&=& 64(q_1^0-{\bf q_1}\cdot\hat{p})(q_2^0-{\bf q_2}\cdot\hat{k})
\nonumber\\
&=& 64 q_1^0 q_2^0 (1-\hat{q}_1\cdot\hat{p})(1-\hat{q}_2\cdot\hat{k})~.
\end{eqnarray}
Contraction between the leptonic $\mathcal{L}^{\mu\nu}(q_1,q_2)$ and the hadronic
 tensor of the anomalous phase $\mathcal{H}_{\mu\nu}^{(\rm a)}(\hat{p},\hat{k})$ leads to
\begin{eqnarray}
\mathcal{L}^{\mu\nu}(q_1,q_2)\mathcal{H}_{\mu\nu}^{(\rm a)}(\hat{p},\hat{k})
&=&  \mathcal{L}^{\mu\nu}(q_1,q_2) \mathcal{T}_{\mu\nu}^-(\hat{p},\hat{k})+ 
\mathcal{L}^{\mu\nu}(q_1,q_2)\mathcal{T}_{\mu\nu}^+(\hat{p},\hat{k})
\nonumber\\
&~&+\mathcal{L}^{\mu\nu}(q_1,q_2)\widetilde{\mathcal{T}}_{\mu\nu}^-(\hat{p},
\hat{k})+\mathcal{L}^{\mu\nu}(q_1,q_2)\widetilde{\mathcal{T}}_{\mu\nu}^+
(\hat{p},\hat{k})
\nonumber\\
&~&-2[\mathcal{L}^{\mu\nu}(q_1,q_2)\mathcal{W}_{\mu\nu}^+(\hat{p},\hat{k})
-\mathcal{L}^{\mu\nu}(q_1,q_2)\mathcal{W}_{\mu\nu}^-(\hat{p},\hat{k})]
\nonumber\\
&=& 64[(q_1^0+{\bf q_1 }\cdot\hat{k})(q_2^0+{\bf q_2}\cdot\hat{p})+(q_1^0
-{\bf q_1 }\cdot\hat{p})(q_2^0-{\bf q_2}\cdot\hat{k})]
\nonumber\\
&=& 64q_1^0q_2^0[(1+\hat{q}_1\cdot\hat{k})(1+\hat{q}_2\cdot\hat{p})+(1
-\hat{q}_1\cdot\hat{p})(1-\hat{q}_2\cdot\hat{k})]~.
\nonumber\\
\end{eqnarray}
\end{widetext}


\end{document}